\documentclass[%
 reprint,
 amsmath,amssymb,
 aps,
]{revtex4-1}

\usepackage{graphicx}% Include figure files
\usepackage{dcolumn}% Align table columns on decimal point
\usepackage{bm}% bold math
\usepackage{float} % for placement of figures
\usepackage{multirow}

\usepackage{subfig}

\usepackage{IEEEtrantools}
\newcommand*{\affaddr}[1]{#1} % No op here. Customize it for different styles.
\newcommand*{\affmark}[1][*]{\textsuperscript{#1}}
\usepackage[USenglish]{babel}
\begin{document}
\bstctlcite{bstctl:nodash}
\bstctlcite{bstctl:etal}
\bstctlcite{bstctl:simpurl}

\preprint{APS/123-QED}

\title{Probing a $\mathrm{Z}^{\prime}$ with non-universal fermion couplings through top quark fusion, decays to bottom quarks, and machine learning techniques} % Force line breaks with \\

\author{
Diego Barbosa\affmark[2], Felipe D\'iaz\affmark[2], Liliana Quintero\affmark[2], Andr\'es Fl\'orez\affmark[2], Manuel Sanchez\affmark[2] \\ Alfredo Gurrola\affmark[1], Elijah Sheridan\affmark[1], Francesco Romeo\affmark[1]\\
\affaddr{\affmark[1] Department of Physics and Astronomy, Vanderbilt University, Nashville, TN, 37235, USA}\\
\affaddr{\affmark[2] Physics Department, Universidad de los Andes, Bogot\'a, Colombia}\\
%\email{\{A,B,C,D,E\}@university.edu}\\
%\affaddr{\LaTeX\ University}
}

\date{\today}% It is always \today, today,

\begin{abstract}

%\begin{center}
% {\bf RESPONSIBLE: Alfredo, Francesco, Andres}
%\end{center}

The production of heavy mass resonances has been widely studied theoretically and experimentally. Several extensions of the standard model (SM) of particle physics, naturally give rise to a new resonance, with neutral electric charge, commonly referred to as the $\textrm{Z}^{\prime}$ boson. The nature, mass, couplings, and associated quantum numbers of this hypothetical particle are yet to be determined. We present a feasibility study on the production of a vector like $\textrm{Z}^{\prime}$ boson at the LHC, with preferential couplings to third generation fermions, considering proton-proton collisions at $\sqrt{s} = 13$ $\mathrm{TeV}$ and 14 TeV. We work under two simplified phenomenological frameworks where the $\mathrm{Z}^{\prime}$ masses and couplings to the SM particles are free parameters, and consider final states of the $\textrm{Z}^{\prime}$ decaying to a pair of $\mathrm{b}$ quarks. The analysis is performed using machine learning techniques in order to maximize the experimental sensitivity. The proposed search methodology can be a key mode for discovery, complementary to the existing search strategies considered in literature, and extends the LHC sensitivity to the $\mathrm{Z}^{\prime}$ parameter space. 
\end{abstract}

\pacs{Valid PACS appear here}% PACS, the Physics and Astronomy
\maketitle

%\tableofcontents

\section{Introduction}

%\begin{center}
% {\bf RESPONSIBLES: Alfredo, Francesco, Andres}
%\end{center}

The standard model (SM) of particle physics is a successful theory to explain a plethora of experimental observations involving weak, electromagnetic, and strong interactions over the last few decades. However, as experiments probe new questions and increasing energies, observations indicate the SM is incomplete and might be a low-energy remnant of a more complete theory. There are a multitude of theoretical models proposed to overcome the SM limitations. Although the initial motivations and resulting implications of these models can vary, a common characteristic is the manifestation of new particles that can be probed in proton-proton (pp) collisions at the CERN's Large Hadron Collider (LHC). 

Numerous ideas have been proposed to probe physics beyond the SM, motivating a large volume of searches at the LHC. Nonetheless, extensive searches have found no firm indication of new phenomena, largely constraining theories and setting exclusion limits up to multi-TeV on the masses of new particles predicted by those theories~\cite{ATLAS:2019fgd, ATLAS:2021jyv, ATLAS:2019isd,CMS:2022zoc,CMS:2021ctt,CMS:2019gwf,CMS:2020wzx,CMS:2022rqc}. Possible explanations for the lack of evidence point to either new particles being too massive or having too low a production rate in existing colliders, or new physics having different features compared to what is traditionally assumed in many beyond SM theories and searches, thus remaining concealed in processes not yet investigated. In particular, many searches conducted so far at the LHC rely heavily on the assumption that these hypothesized new particles have similar couplings to all generations of fermions, including couplings to the partons inside the proton, thus favoring LHC production modes through light quarks. Therefore, if new phenomena are within the reach of the LHC, both in energy and production rate, they might manifest with different features compared to what is assumed in searches at high energy colliders, thus requiring new efforts and experimental quests.

In this paper, we consider a different scenario in which new particles have non-universal fermion couplings, favoring higher-generation fermions, which we refer to as $anogenophilic$ particles. In particular, we consider a new neutral vector gauge boson, $\textrm{Z}^{\prime}$, with only couplings to third generation fermions, referred to as $tritogenophilic$. This physics case is also interesting theoretically and because of recent results in precision measurements, offering a new physics phase space not yet fully explored at the LHC.

An anogenophilic resonance is predicted in several theories that extend the SM and in different contexts such as in top-assisted technicolor models~\cite{Hill:1994hp}, Randall–Sundrum models with Kaluza–Klein excitations of the graviton~\cite{Randall:1999ee, Randall:1999vf,Davoudiasl:2000wi}, two-Higgs doublet models that address the naturalness of the electroweak symmetry breaking scale~\cite{Branco:2011iw,Gori:2016zto,BhupalDev:2014bir,Gunion:1989we}, left-right extensions of the SM~\cite{left-right}, models with a color-sextet or color-octet~\cite{Dobrescu:2007yp,Chen:2008hh,Bai:2010dj,Berger:2010fy,Zhang:2010kr,Gerbush:2007fe}, with composite particles~\cite{ArkaniHamed:2002qy, Pomarol:2008bh,Matsedonskyi:2012ym,Gripaios:2014pqa,Liu:2015hxi,Cacciapaglia:2015eqa,Barbieri:2007bh,Panico:2011pw,DeCurtis:2011yx,Marzocca:2012zn,Bellazzini:2012tv,Panico:2012uw,Bellazzini:2014yua,Compositeness_GurrolaRomeo}, or with dark matter mediators~\cite{Haisch:2015ioa,Buckley:2014fba,Cox:2015afa,Zhang:2012da,Baek:2016lnv,Arina:2016cqj,DHondt:2015nat,Baek:2017sew}. 
Furthermore, such a pattern pointing to non-universal fermion couplings appears in several intriguing experimental results from precision measurements in the $B$-physics sector~\cite{Lees:2012xj,Lees:2013uzd,Huschle:2015rga,Sato:2016svk,Hirose:2016wfn,Hirose:2017dxl, Belle:2019rba, Aaij:2015yra,Aaij:2017uff,Aaij:2017tyk,Wehle:2016yoi,Aaij:2013qta,Aaij:2014pli,Aaij:2014ora,Aaij:2015oid,Aaij:2015esa,Aaij:2017vbb,Aaij:2019wad} and the measurement of the muon anomalous magnetic moment~\cite{Muong-2:2021ojo}, all showing significant deviations from the SM expectation~\cite{Altmannshofer:2021qrr,HFAG,Muong-2:2021ojo}. 
%, despite yet to be confirmed. 
The scenario of a tritogenophilic new particle is also interesting in light of recent results of the measured cross sections for pp$\to\mathrm{t} \bar{\mathrm{t}} + \mathrm{b} \bar{\mathrm{b}}$ ($\mathrm{t} \bar{\mathrm{t}} \mathrm{t} \bar{\mathrm{t}}$) production from the ATLAS~\cite{ATLAS:2018fwl,ATLAS:2020hpj} (\cite{ATLAS:2021kqb}) and CMS~\cite{CMS:2020grm,CMS:2019eih} collaborations, which are found to be higher than the expectations from the SM.

The mass, quantum numbers, and couplings of new hypothetical mediators can be open parameters to be determined experimentally, making the new physics phase space broadly defined. Thus, initial ATLAS/CMS searches for these new type of particles were conducted considering models with democratic couplings to all fermion families, and focused on Drell-Yan production mechanisms with light quarks (e.g., $\mathrm{q\bar{q}}\to\mathrm{Z}^{\prime}$), and final states with muons and electrons with high signal acceptance and a narrow ``bump’’ in the reconstructed invariant mass spectrum of lepton pairs sitting above a smooth and steeply falling background distribution \cite{ATLAS:2019erb, CMS:2019tbu}. However, from the phenomenological point of view, when couplings to light quarks are suppressed in pp colliders, relative to higher-generation fermions, new production mechanisms become dominant to generate and discover beyond SM resonances. They are produced in association with other SM particles and give origin to rare and peculiar signatures.
The phenomenology of purely top-philic $\textrm{Z}^{\prime}$~\cite{Greiner:2014qna, Kim:2016plm, Fox:2018ldq} scenarios, as well as models with a $\textrm{Z}^{\prime}$ that couples to top quarks and tau/muon~\cite{Abdullah:2019dpu,Kamenik:2017tnu} leptons, have already been studied in the literature. Furthermore, a CMS search has been performed for a neutral resonance coupling to top quarks and decaying to muons or electrons~\cite{CMS:2019lwf}.

\begin{figure}
 \begin{center} 
  \includegraphics[width=0.4\textwidth]{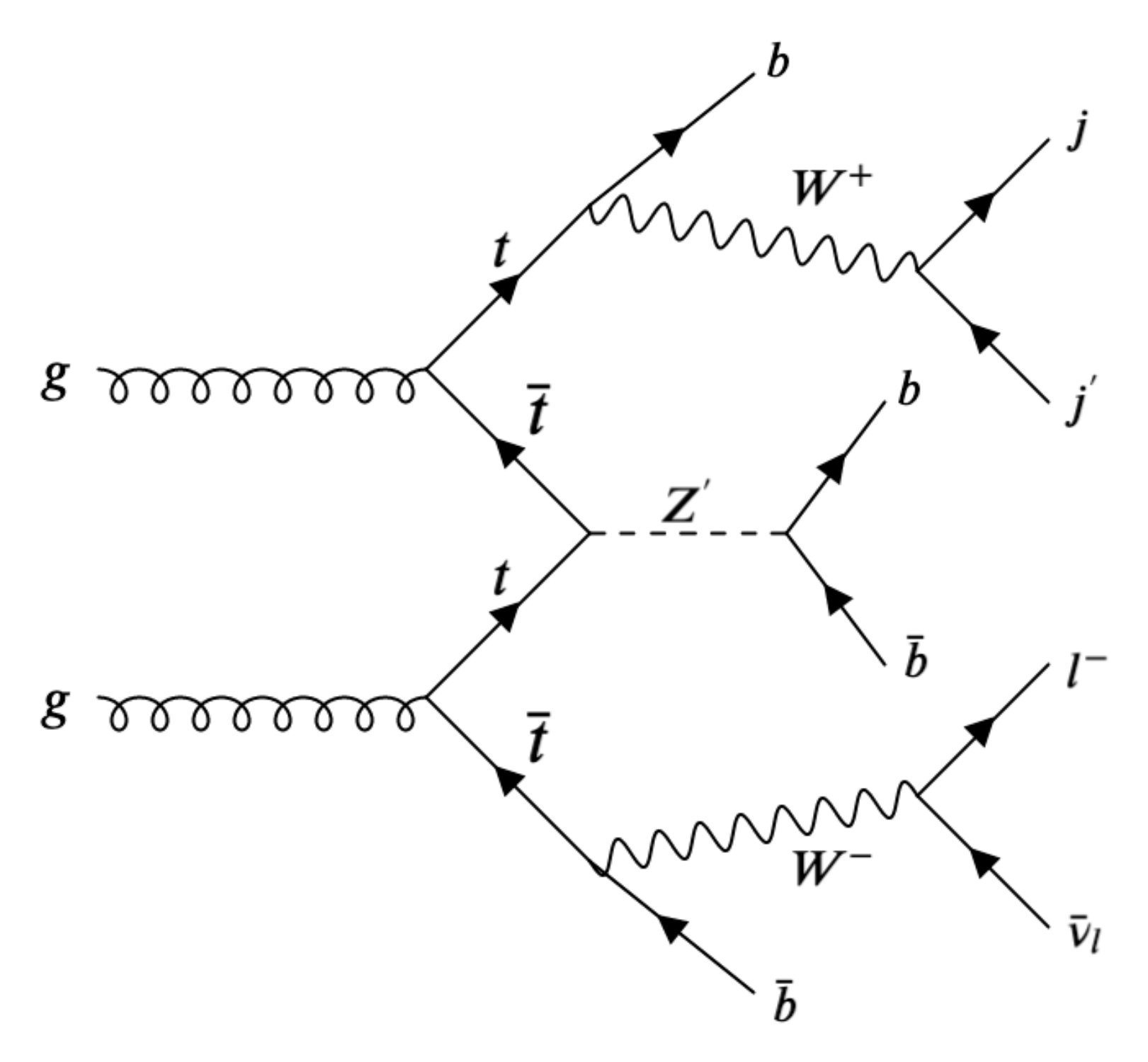}
 \end{center}
  \vspace{-0.7cm}
 \caption{Representative Feynman diagram for the production of a $\textrm{Z}^{\prime}$ boson through the fusion of a top quark pair, where the $\textrm{Z}^{\prime}$ decays to a pair of bottom quarks and the two spectator top quarks decay semi-leptonically.}
 \label{SMprocessNewPhysics}
 \end{figure}

In this paper, we perform a previously unexamined feasibility study on the production of a more general tritogenophilic $\textrm{Z}^{\prime}$ produced through the fusion of a $\mathrm{t} \bar{\mathrm{t}}$ pair ($\mathrm{t} \bar{\mathrm{t}} \textrm{Z}^{\prime}$) and decaying to a pair of $\mathrm{b}$ quarks ($\textrm{Z}^{\prime}\rightarrow \mathrm{b}\bar{\mathrm{b}}$), as in Figure~\ref{SMprocessNewPhysics}. We consider the final state where one of the two remaining tops from the fusion process, referred to as $spectator$ top quarks, decays to $\mathrm{b}\mathrm{W}$ and the $\mathrm{W}$ boson subsequenly decays to an electron or muon plus its neutrino. Such a choice balances the lower $\mathrm{W}\to \ell\nu$ branching fraction compared to $\mathrm{W}$ boson decays into two quarks, with a cleaner final state. This has the double advantage of mitigating the large background from full-hadronic SM quantum chromodynamics (QCD) processes, and of overcoming the otherwise overwhelming events rate that is outside the typical trigger bandwidth at the LHC, rendering the search sensitive to a wide range of $\textrm{Z}^{\prime}$ masses. For $\textrm{Z}^{\prime}$ masses below the $2m_{\mathrm{t}}$ kinematic production threshold, where $\textrm{Z}^{\prime} \to \mathrm{t}\bar{\mathrm{t}}$ decays are not permitted, the $\textrm{Z}^{\prime}$ decay to $\mathrm{b} \bar{\mathrm{b}}$ is the dominant discovery mode. Furthermore, the analysis strategy proposed in this paper provides enhanced sensitivity compared to other approaches already used in searches at the LHC~\cite{ATLAS:2019itm, ATLAS:2019fgd, CMS:2022zoc, CMS:2018kcg, CMS:2018pwl}.
Above $2m_{\mathrm{t}}$, the reduced jet multiplicity of the $\textrm{Z}^{\prime} \to \mathrm{b}\bar{\mathrm{b}}$ final state, in comparison to $\textrm{Z}^{\prime} \to \mathrm{t}\bar{\mathrm{t}}$, favors the experimental reconstruction of the $\textrm{Z}^{\prime}$ mass. In this work, machine learning techniques are used to maximize the experimental sensitivity.

\section{Samples and simulation}

Signal and background samples are generated with MadGraph5\_aMC (v2.6.3.2)~\cite{Alwall:2014hca} considering  $\textrm{pp}$ beams colliding with a center-of-mass energy of $\sqrt{s}=13$ TeV and $\sqrt{s}=14$ $\mathrm{TeV}$. All samples are generated using the NNPDF3.0 NLO~\cite{NNPDF:2014otw} set for parton distribution functions (PDFs). Parton level events are then interfaced with the PYTHIA (v8.2.05)~\cite{Sjostrand:2014zea} package to include parton fragmentation and hadronization processes, while DELPHES (v3.4.1)~\cite{deFavereau:2013fsa} is used to simulate detector effects, using the CMS detector geometric configurations and parameters, for performance of particle reconstruction and identification. 
At parton level, jets are required to have a minimum transverse momentum ($p_{T}$) of 20 $\mathrm{GeV}$ and pseudorapidity ($\eta$) $|\eta| < 5.0$. The cross sections in this paper are obtained with the aforementioned parton-level selections. The MLM algorithm~\cite{Alwall:2007fs} is used for jet matching and jet merging. The xqcut and qcut variables of the MLM algorithm, related with the minimal distance between partons and the energy spread of the clustered jets, are set to 30 and 45, respectively, as a result of an optimization process requiring the continuity of the differential jet rate as a function of jet multiplicity. 

The signal samples are generated considering the production of a $\mathrm{Z}^{\prime}$ and two associated top quarks ($\mathrm{pp}\to\mathrm{Z}^{\prime}\mathrm{t}\bar{\mathrm{t}}$), inclusive in  $\alpha_{\textrm{EWK}}$ and $\alpha_{\textrm{QCD}}$. 
For our benchmark signal scenario, we consider the simplified model in Ref.~\cite{VBFZprimePaper} where the $\mathrm{Z}^{\prime}$ masses and couplings to the SM particles are free parameters, and defined as variations of the SM $\mathrm{Z}$ boson couplings (i.e., variations of the so-called Sequential Standard Model, SeqSM). The $\mathrm{Z}^{\prime}$ coupling to the first and second generation SM quarks is defined as $g_{\mathrm{Z}^{\prime}q\bar{q}} = g_{q} \times g_{\mathrm{Z}q\bar{q}}$, where $g_{\mathrm{Z}q\bar{q}}$ is the SM $\mathrm{Z}$ boson coupling to first and second generation quarks and $g_{q}$ is a ``modifier'' for the coupling. Similarly, the $\mathrm{Z}^{\prime}$ coupling to the third generation SM quarks is defined as $g_{\mathrm{Z}^{\prime},\mathrm{b}/\mathrm{t},\bar{\mathrm{b}}/\bar{\mathrm{t}}} \times g_{ \mathrm{Z},\mathrm{b}/\mathrm{t},\bar{\mathrm{b}}/\bar{\mathrm{t}}}$, where $g_{\mathrm{Z}^{\prime},\mathrm{b}/\mathrm{t},\bar{\mathrm{b}}/\bar{\mathrm{t}}}$ is the modifier to the SeqSM coupling. We refer to this model as ``simplified phenomenological model 1'' (SPM1). In all cases considered, the modifiers for the $\mathrm{Z}^{\prime}$ couplings to $\mathrm{t}\bar{\mathrm{t}}$ and $\mathrm{b}\bar{\mathrm{b}}$ are equal to each other, and thus for simplicity we henceforth refer to those modifiers as $g_{\mathrm{Z}^{\prime}\mathrm{t}\bar{\mathrm{t}}}$. Therefore, a scenario with $g_{\mathrm{Z}^{\prime}\mathrm{t}\bar{\mathrm{t}}} = 1$ has similar $\mathrm{Z}^{\prime}$ couplings to top/bottom quarks as the SeqSM. Signal samples were created for $m(\mathrm{Z}^{\prime})$ ranging from 250 GeV to 2000 GeV. Table~\ref{tabla_cs} lists the production cross sections for different  $\textrm{Z}^{\prime}$ masses, considering $\textrm{pp}$ collisions at $\sqrt{s}=13$ $\textrm{TeV}$ and 14 $\textrm{TeV}$, and for two representative $g_{q}$ coupling scenarios with $g_{\mathrm{Z}^{\prime}\mathrm{t}\bar{\mathrm{t}}} = 1$. The $g_{q} = 0$ case is a proxy for the tritogenophilic scenarios, where the couplings of the $\mathrm{Z}^{\prime}$ to light quarks are suppressed. The $g_{q} = 1$ case allows for non-negligible couplings to light quarks, and thus other $\mathrm{t\bar{t}}\mathrm{Z}^{\prime}$ production processes can contribute, such as initial state radiation of a $\mathrm{Z}^{\prime}$ from a light quark.

In addition to our primary signal benchmark model described above, we also consider a tritogenophilic scenario where the $\mathrm{Z}^{\prime}$ is a color singlet vector particle whose effective couplings are not suppressed by factors of the electroweak mixing angles (as in the SeqSM) and whose relevant interactions to top/bottom quarks are given by the following renormalizable Lagrangian: $\mathcal{L}_{\mathrm{int}} = \bar{\mathrm{t}}\gamma_{\mu}(c_{L}P_{L}+c_{R}P_{R})\mathrm{t}\mathrm{Z}^{\prime\mathrm{ }\mu} = c_{\mathrm{eff}}\bar{\mathrm{t}}\gamma_{\mu}(\mathrm{cos}\theta P_{L} + \mathrm{sin}\theta P_{R})\mathrm{t}\mathrm{Z}^{\prime\mathrm{ }\mu}$, where $P_{R/L} = (1 \pm \gamma_{5})/2$ are the projection operators, $c_{\mathrm{eff}} = \sqrt{c_{L}^{2} + c_{R}^{2}}$ is the $\mathrm{Z}^{\prime}$ coupling to top/bottom quarks, and $\mathrm{tan}\theta = c_{R}/c_{L}$ is the tangent of the chirality angle. We consider the case where the $\mathrm{Z}^{\prime}$ couplings to top and bottom quarks are equal to each other, and thus for simplicity
we henceforth refer to those couplings as $c_{\mathrm{t}}$. This type of simplified model, which we refer to as ``simplified phenomenological model 2'' (SPM2), has been studied in Refs.~\cite{Greiner:2014qna, Kim:2016plm, Fox:2018ldq}, and it has been shown that $\mathrm{t}\bar{\mathrm{t}}\mathrm{Z}^{\prime}$ production is independent of $\theta$. We have checked that this indeed the case. Thus, we only consider $\theta = \pi/2$. Although the signal kinematic distributions for this particular model are similar to those of SPM1, the $\mathrm{t}\bar{\mathrm{t}}\mathrm{Z}^{\prime}$ production cross sections for SPM2 are larger than those of SPM1, when $c_{\mathrm{t}} = g_{\mathrm{Z}^{\prime}\mathrm{t\bar{t}}}$, since the SPM2 Lagrangian does not contain suppression terms from the electroweak mixing angles. Our primary motivation in using SPM2 is to compare the projected discovery reach of the proposed analysis strategy in this paper, with other strategies, such as those in Ref.~\cite{Kim:2016plm}, which considers the $\mathrm{Z}^{\prime}\to\mathrm{t\bar{t}}$ decay mode.

Several sources of background are considered for our studies, including production of top quark pairs ($\mathrm{t}\bar{\mathrm{t}}$), $\textrm{Z/W}$ bosons with associated jets ($\textrm{V}+$jets), QCD multijet, associated production of a Higgs ($\mathrm{h}$) or a $\mathrm{Z/\gamma^{*}}$ boson from $\mathrm{t}\bar{\mathrm{t}}$ fusion processes (denoted $\mathrm{t}\bar{\mathrm{t}} \mathrm{h}$ and $\mathrm{t}\bar{\mathrm{t}} \mathrm{X}$), and associated production of four $\mathrm{t}$ quarks ($\mathrm{t}\bar{\mathrm{t}}\mathrm{t}\bar{\mathrm{t}}$). 
Since our signal topology targets final states with four bottom quarks ($\mathrm{Z}^{\prime}\to\mathrm{b\bar{b}}$ and $\mathrm{t\bar{t}}\to\mathrm{bWbW}$), the $\mathrm{t}\bar{\mathrm{t}}$, $\textrm{V}+$jets, and QCD multijet backgrounds do not meaningfully contribute to our studies ($\ll 1$\% of the total background). The $\mathrm{t}\bar{\mathrm{t}}\mathrm{h}$, $\mathrm{t}\bar{\mathrm{t}}\mathrm{X}$, and $\mathrm{t}\bar{\mathrm{t}}\mathrm{t}\bar{\mathrm{t}}$ processes are the dominant sources of background events. The $\mathrm{t\bar{t}h}$ and $\mathrm{t\bar{t}X}$ processes become important backgrounds when $\mathrm{h}$ and $\mathrm{Z}/\gamma^{*}$ decay to a pair of bottom quarks. Table~\ref{tabla_csbkg} shows the production cross sections for the dominant backgrounds, at $\sqrt{s}=13$ $\textrm{TeV}$ and 14 $\textrm{TeV}$.

\begin{table}[ht!]
    \centering
 \caption{Signal cross sections, calculated with MadGraph, for different $\textrm{Z}^{\prime}$ masses and couplings to first and second generation quarks. The values in this table are calculated with $g_{\mathrm{Z}^{\prime}\mathrm{t\bar{t}}} = 1$.}
    \begin{tabular}{c|c|c|c|c}
    \hline
    %\hline
    \multicolumn{1}{c|}{}&\multicolumn{2}{c|}{13 $\textrm{TeV}$}&\multicolumn{2}{c}{14 $\textrm{TeV}$}\\
    \hline
    $\textrm{Z}^{\prime}$ mass ($\mathrm{GeV}$) & $\sigma_{g_{q} = 0}$($\textrm{fb}$) & $\sigma_{g_{q} = 1}$($\textrm{fb}$) & $\sigma_{g_{q} = 0}$($\textrm{fb}$) & $\sigma_{g_{q} = 1}$($\textrm{fb}$)\\
    \hline

    250  & 51.34   & 72.87  & 64.32  & 90.21\\
    300  & 32.61   & 47.73  & 41.22  & 58.94\\
    325  & 26.40   & 39.47  & 33.48  & 48.33\\
    350  & 21.58   & 32.84  & 27.14  & 40.53\\
    375  & 17.71   & 27.48  & 22.50  & 33.77\\
    400  & 14.58   & 23.14  & 18.97  & 28.62\\
    500  & 7.379   & 12.53  & 9.557  & 15.67\\
    750  & 1.700   & 3.546  & 2.315  & 4.516\\
    1000 &0.502   & 1.285  & 0.703  & 1.681\\
    2000 &0.011   & 0.066  & 0.017  & 0.093\\
    
    \hline
    \hline
    \end{tabular}
    \label{tabla_cs}
\end{table}

\begin{table}[ht!]
    \centering
 \caption{Cross sections calculated with MadGraph for the dominant background processes.}
    \begin{tabular}{c|c|c}
    \hline
    
    \hline
    Process & 13 $\textrm{TeV}$ ($\textrm{fb}$) & 14 $\textrm{TeV}$ ($\textrm{fb}$)\\
    \hline
    $t\Bar{t} \mathrm{h}$     & 393.5   &  476.3 \\
    $t\Bar{t} \mathrm{X}$     & 13600   & 16620  \\
    $t\Bar{t}t\Bar{t}$ & 8.973   & 11.80  \\

    \hline
    \hline
    \end{tabular}
    \label{tabla_csbkg}
\end{table}

The total event rates are determined using $N = \sigma \times \textrm{L} \times \epsilon$, where $N$ represents the total yield of events, $\textrm{L}$ the integrated luminosity considered (for this study, 150 fb$^{-1}$, 300 fb$^{-1}$, and 3000 fb$^{-1}$), and $\epsilon$ represents any efficiencies which might reduce the total event yield (e.g., particle identification efficiencies). The $\textrm{L} = 150$ fb$^{-1}$ scenario represents an estimate for the amount of data already collected by the ATLAS and CMS experiments, while the other luminosity scenarios are the expectations for the next decade of $\textrm{pp}$ data taking at the LHC. All production cross sections are computed at tree-level. Since the k-factors associated with higher-order corrections to QCD production cross sections are typically greater than one, our estimates of the sensitivity are conservative.

Following Ref.~\cite{CMS_BTV2016}, we consider three possible ``working points'' for the identification of the b-jet candidates in DELPHES: (i) the ``Loose'' working point of the DeepCSV algorithm, which gives a 85\% b-tagging efficiency and 10\% light quark mis-identification rate; (ii) the ``Medium'' working point of the DeepCSV algorithm, which gives a 70\% b-tagging
efficiency and 1\% light quark mis-identification rate; and (iii) the ``Tight'' working point of the DeepCSV algorithm, which gives a 45\% b-tagging efficiency and 0.1\% light quark mis-identification rate. The choice of $\mathrm{b}$-tagging working points is determined through an optimization process which maximizes discovery reach. The ``Medium'' working point was ultimately shown to provide the best sensitivity and therefore chosen for this study. For muons (electrons), the assumed identification efficiency is 95\% (85\%), with a 0.3\% (0.6\%) mis-identification rate~\cite{CMS-PAS-FTR-13-014,CMS_MUON_17001,CMS_EGM_17001}.

\section{DATA ANALYSIS USING THE GRADIENT BOOST ALGORITHM}

The analysis of signal and background events is performed using a machine learning event classifier, namely a gradient boosted decision trees (BDTs) \cite{friedman_greedy_2001}. Machine learning offers advantages over traditional event classification methods. %, such as iteratively optimizing event selection criteria one at a time. 
In particular, machine learning models consider all kinematic variables in tandem, efficiently traversing the high-dimensional space of event kinematics, thereby enabling them to enact complicated selection criteria which incorporates that high-dimensional space in its entirety.

This method iteratively trains decision trees to learn the residuals between predictions and expected values yielded by the tree trained just before it, thereby greedily minimizing error at each iteration. BDTs have been employed to great effect previously in classification problems arising in collider physics (e.g., \cite{Ai:2022qvs,ATLAS:2017fak,Chigusa:2022svv,Chung:2020ysf,Feng:2021eke,CMS:2014tll,Santos:2016kno}).  

Simulated signal and background events are initially filtered, before being passed to the BDT algorithm, requiring at least four well reconstructed and identified $\mathrm{b}$-jet candidates, at least two jets not tagged as $\mathrm{b}$ jets, and exactly one identified light lepton ($\ell$), that could be either an electron $(\mathrm{e})$ or a muon ($\mu$). Selected jets must have $p_{\mathrm{T}} > 30$ $\textrm{GeV}$ and $|\eta(j)| < 5.0$, while $\mathrm{b}$-jet candidates with $p_{\mathrm{T}} > 30$ $\textrm{GeV}$ and $|\eta(\mathrm{b})| < 2.5$ are chosen. The $\ell$ object must pass a $p_{\mathrm{T}} > 25$ $\textrm{GeV}$ threshold and be within a $|\eta(\ell)| < 2.5$. Overlapping objects in $\eta-\phi$ space are removed using a minimum $\Delta R$ among all particle candidates ($p_{i}$) above 0.3, where $\Delta R (p_{i}, p_{j}) = \sqrt{ (\Delta \phi (p_{i}, p_{j}))^{2} + (\Delta \eta (p_{i}, p_{j}) )^{2} }$. These filtering criteria will be henceforth referred to as pre-selections. Table \ref{tab:selection_criteria} summarizes these pre-selections for the analysis. 

Events passing this pre-selection are used as input for the BDT algorithm, which classifies them as signal or background, using a probability factor. We implement the BDT algorithm using the canonical \texttt{scikit-learn} \cite{pedregosa_scikit-learn_2011} and \texttt{xgboost} \cite{chen_xgboost_2016} libraries. In particular, we employed the \texttt{XGBClassifier} class in the latter library with $250$ iterations, a max depth of $7$, a learning rate of $0.1$, and default parameters otherwise, although we note that model performance was found to be largely independent of hyperparameters.

\begin{figure}[]
    \centering
    \includegraphics[width=0.48\textwidth]{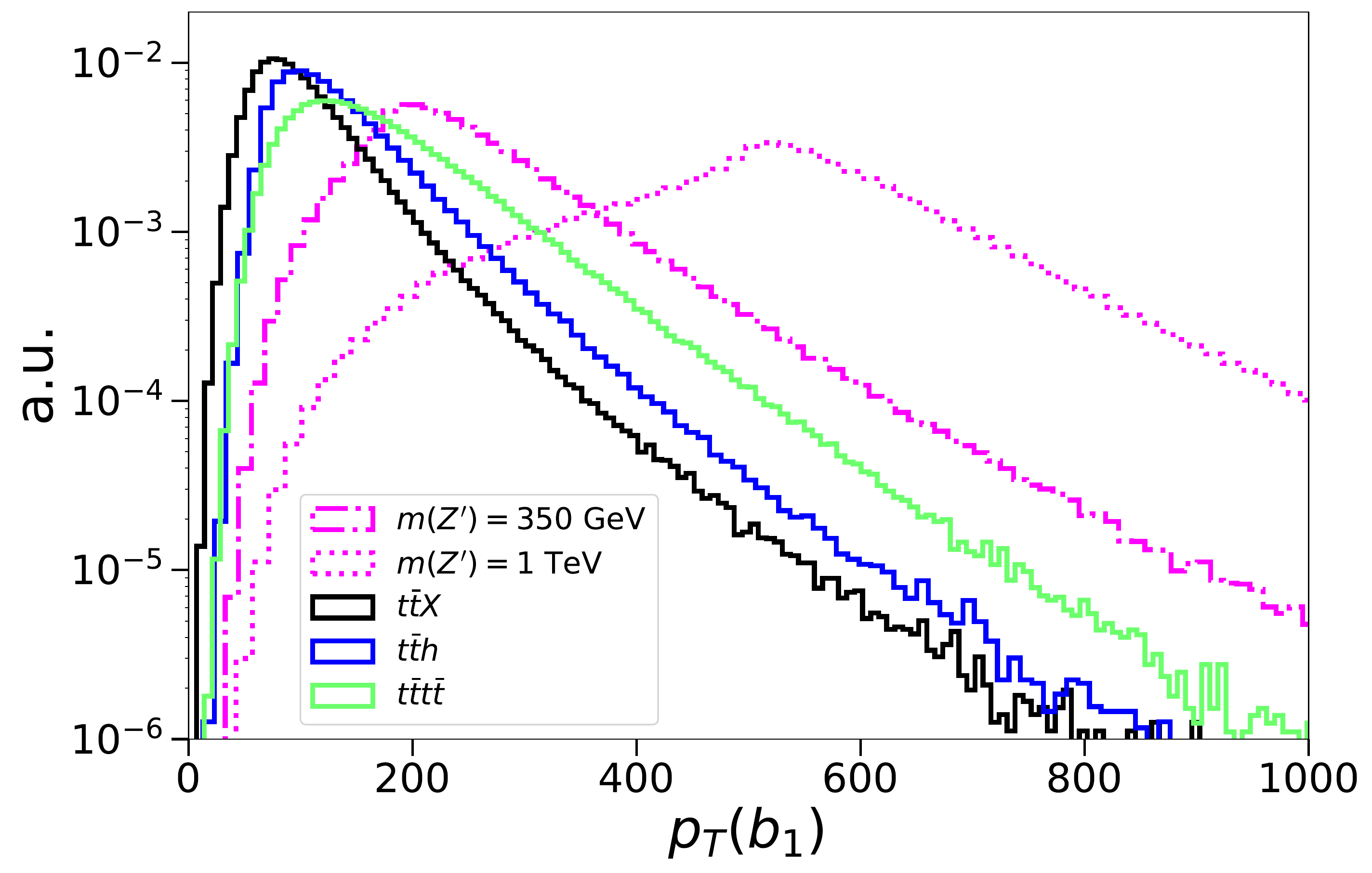}
    \caption{Transverse momentum distributions for the $\mathrm{b}$ quark jet with the highest transverse momentum, for two signal points with masses of 350  $\mathrm{GeV}$ and 1000 $\mathrm{GeV}$ and dominant backgrounds.}
    \label{fig:ptb1}
\end{figure}

\begin{figure}[]
    \centering
    \includegraphics[width=0.48\textwidth]{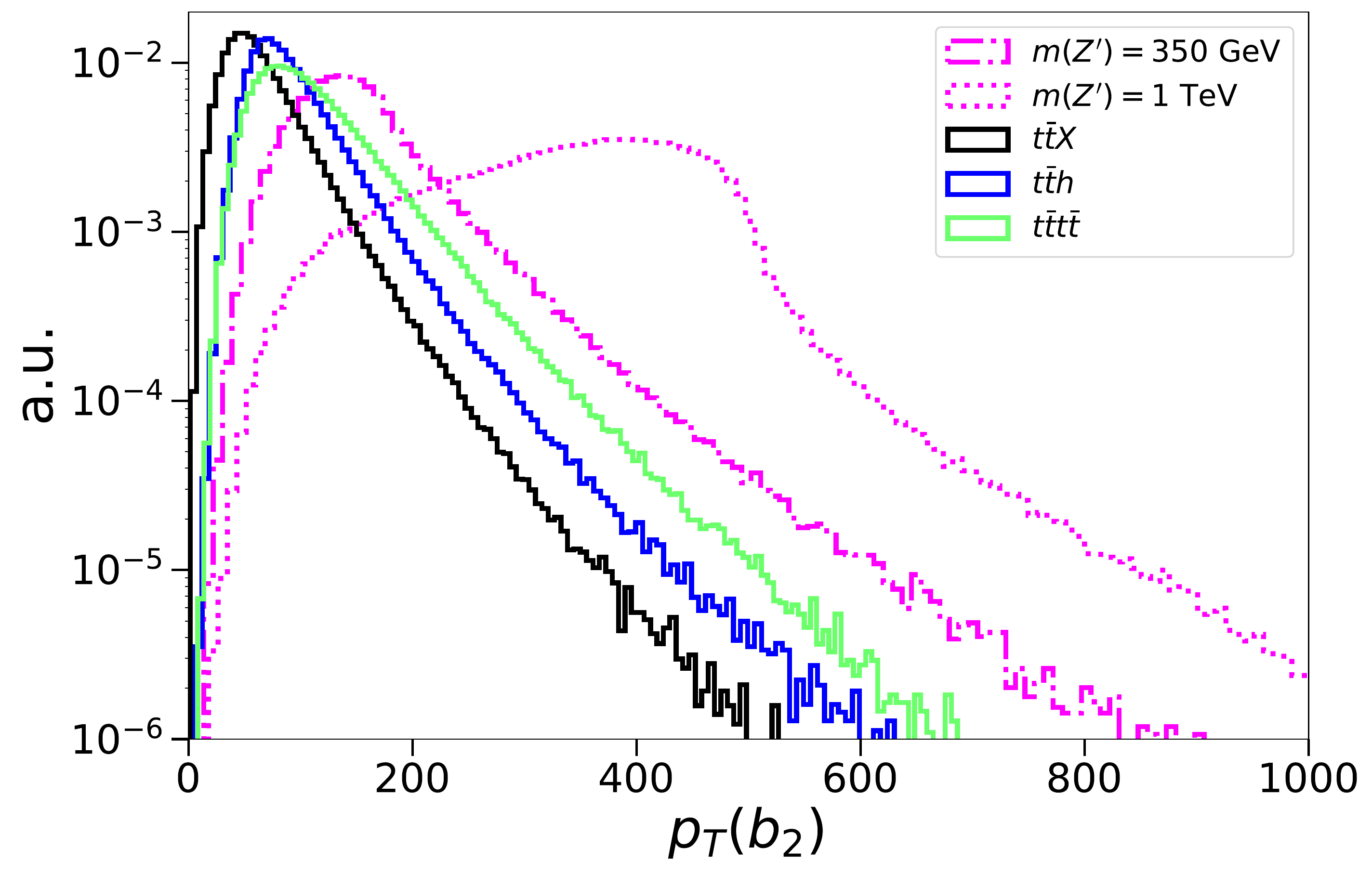}
    \caption{Transverse momentum distributions for the $\mathrm{b}$ quark jet with the second highest transverse momentum, for two signal points with masses of 350  $\mathrm{GeV}$ and 1000 $\mathrm{GeV}$ and dominant backgrounds.}
    \label{fig:ptb2}
\end{figure}

\begin{figure}[]
    \centering
    \includegraphics[width=0.48\textwidth]{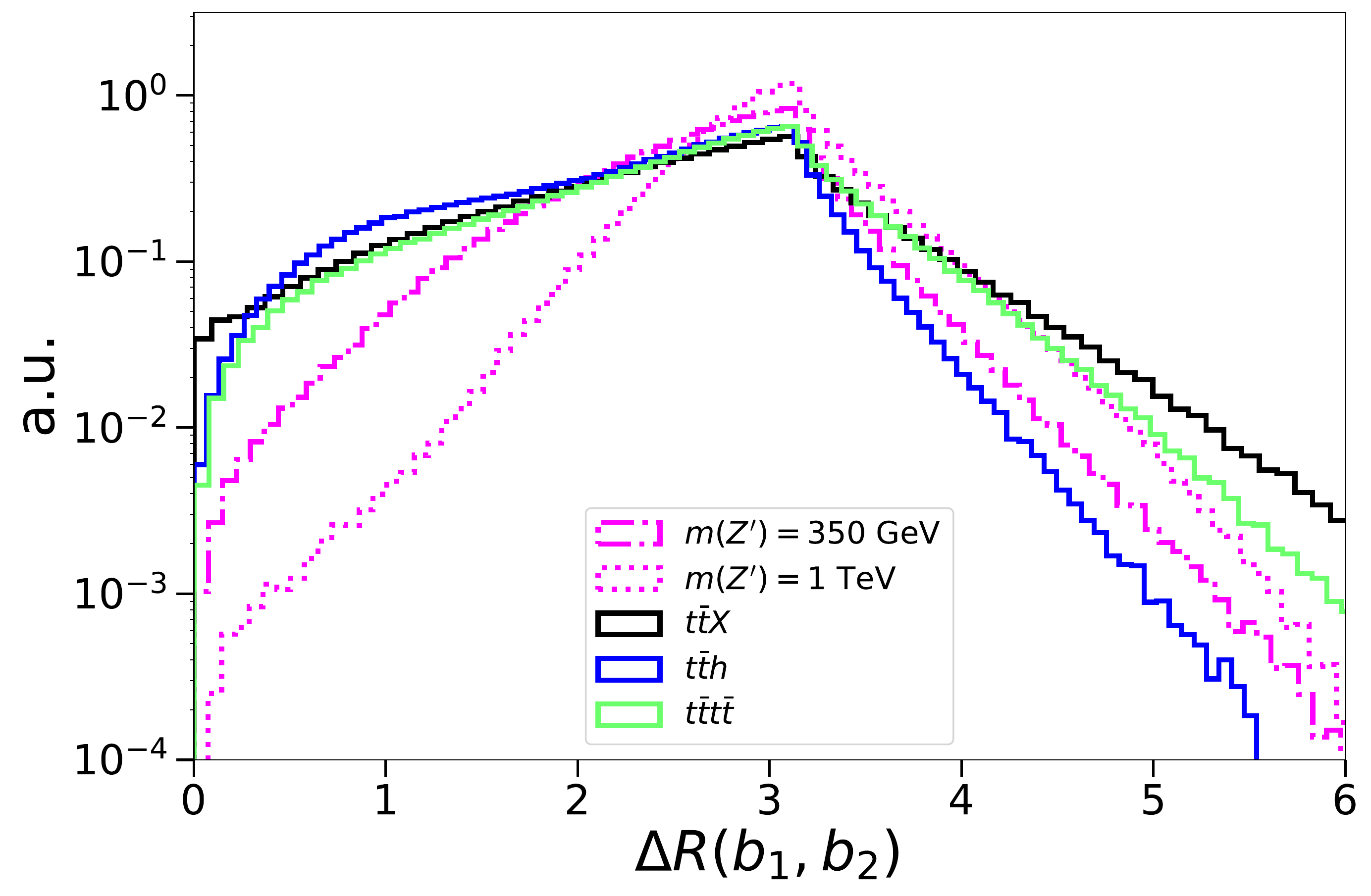}
    \caption{Distributions for the $\Delta R$ angular separation between the the highest ($\mathrm{b_{1}}$) and second highest ($\mathrm{b_{2}}$) transverse momentum $\mathrm{b}$ quark pair, for two signal points with masses of 350  $\mathrm{GeV}$ and 1000 $\mathrm{GeV}$  and dominant backgrounds.}
    \label{fig:dRb1b2}
\end{figure}

\begin{figure}[]
    \centering
    \includegraphics[width=0.48\textwidth]{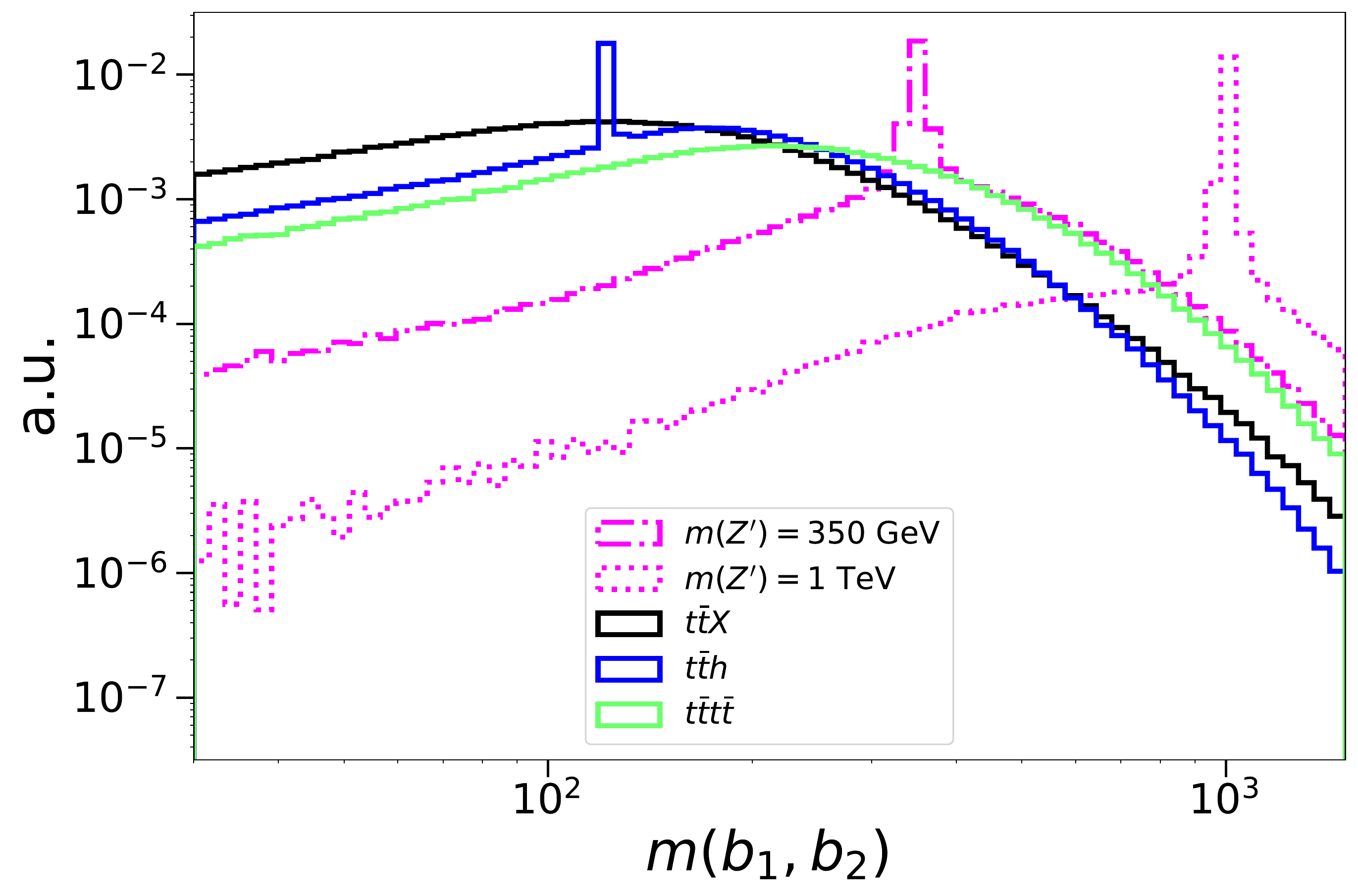}
    \caption{Invariant mass distributions for the highest ($\mathrm{b_{1}}$) and second highest ($\mathrm{b_{2}}$) transverse momentum $\mathrm{b}$ quark pair, for two signal points with masses of 350  $\mathrm{GeV}$ and 1000 $\mathrm{GeV}$ and dominant backgrounds.}
    \label{fig:mb1b2}
\end{figure}

Figures \ref{fig:ptb1}, \ref{fig:ptb2}, \ref{fig:dRb1b2}, and \ref{fig:mb1b2}, show relevant kinematic distributions for two SPM1 signal points and dominant backgrounds, normalized to the area under the curve (unity). The distributions correspond to the $\mathrm{b}$-jet candidate with the highest $p_{\mathrm{T}}$ ($\mathrm{b_{1}}$), the second $\mathrm{b}$-jet candidate with the highest $p_{\mathrm{T}}$ ($\mathrm{b_{2}}$), the $\Delta R$ separation between  the $\mathrm{b_{1}}$ and $\mathrm{b_{2}}$ candidates, and the reconstructed mass between the $\mathrm{b_{1}}$ and $\mathrm{b_{2}}$, $m(\mathrm{b}_{1}, \mathrm{b}_{2})$, respectively. These distributions are among the variables identified by the BDT algorithm with the highest signal to background discrimination power.   

As can be seen from Figures 2 and 3, for $m(\mathrm{Z}^{\prime})$ values beyond the electroweak scale, the relatively large leading and subleading $\mathrm{b}$-jet $p_{\textrm{T}}$ is a key
feature attributed to the heavy $\mathrm{Z}^{\prime}$ with respect to the mass of the bottom quarks, thus resulting in an average $p_{\textrm{T}}(\mathrm{b}_{1,2})$ of approximately $m(\mathrm{Z}^{\prime})/2$. This kinematic feature provides a nice handle to discriminate high $m(\mathrm{Z}^{\prime})$ signal events amongst the large SM backgrounds, which have lower average $p_{\textrm{T}}(\mathrm{b}_{1,2})$ constrained by the top quark and/or higgs masses. The $\Delta R$ separation between $\mathrm{b}_{1}$ and $\mathrm{b}_{2}$ is determined by the amount of momentum transfer to the resonant particles in each process ($\mathrm{Z}^{\prime}$, $\mathrm{h}$, or $\mathrm{t}$), which in turn depends on the masses of those particles. Therefore, Figure 4 shows greater discrimination between background and signal processes as $m(\mathrm{Z}^{\prime})$ becomes larger. Finally, as noted previously, an advantage of the $\mathrm{Z}^{\prime}\to\mathrm{b}\bar{\mathrm{b}}$ final state in comparison to $\mathrm{Z}^{\prime}\to\mathrm{t}\bar{\mathrm{t}}$ is the experimental reconstruction of the $\mathrm{Z}^{\prime}$ mass, which is observed as a peak in the $m(\mathrm{b}_{1}, \mathrm{b}_{2})$ signal distributions in Figure 5 near the true $m(\mathrm{Z}^{\prime})$ value. On the other hand, the background $m(\mathrm{b}_{1}, \mathrm{b}_{2})$ distributions show a peak near $m(\mathrm{h}) = 125$ GeV for the $\mathrm{t}\bar{\mathrm{t}}\mathrm{h}$ background, or a broad distribution for the other backgrounds, indicative of the combination of two b jets from different decay vertices. We note that the $\mathrm{Z}^{\prime}\to\mathrm{b}\bar{\mathrm{b}}$ decay width depends on $g_{\mathrm{Z}^{\prime}\mathrm{t}\bar{\mathrm{t}}}^{2}\times\frac{m_{\mathrm{b}}^{2}}{m(\mathrm{Z}^{\prime})^{2}}$ and is thus suppressed by the relatively small bottom quark mass with respect to the  $g_{\mathrm{Z}^{\prime}\mathrm{t}\bar{\mathrm{t}}}$ and $m(\mathrm{Z}^{\prime})$ values considered in these studies. Therefore, the width of the $m(\mathrm{b}_{1}, \mathrm{b}_{2})$ signal distributions is driven by the experimental resolution in the reconstruction of the $\mathrm{b}$-jet momenta, as well as the probability that the two leading $\mathrm{b}$ jets are the correct pair from the $\mathrm{Z}^{\prime}$ decay. 

In addition to these aforementioned variables in Figures \ref{fig:ptb1}-\ref{fig:mb1b2}, a variety of other kinematic variables were included 
as inputs to the BDT algorithm. In particular, 47 such variables were used in total, 
and these included the momenta of $\mathrm{b}$ and light quark jets  (not tagged as $\mathrm{b}$ jets); invariant masses of pairs of $\mathrm{b}$ jets and of the two leading light jets; angular differences between 
$\mathrm{b}$ jets, between light quark jets, and between the lepton and $\mathrm{b}$ jets; and 
transverse masses derived from the lepton-$p^{miss}_{\mathrm{T}}$ pair and lepton-$p^{miss}_{\mathrm{T}}$-$\mathrm{b}$ triplets. 
The variables $m(\mathrm{b}_{i}, \mathrm{b}_{j})$ for $i, j \neq 1$ 
provide some additional discrimination between signal and background when the leading $\mathrm{b}$-jets are not a $\mathrm{Z}^{\prime}$ decay candidate. The transverse mass 
variables are designed to be sensitive to a leptonic decay of the $\mathrm{W}$ boson and $\mathrm{t}$ quark (i.e., $m_{jj}$ and $m_{\mathrm{T}}(\ell,p_{\mathrm{T}}^{miss})$ should be near $m_{\mathrm{W}}$, and $m_{\mathrm{T}}(\ell,\mathrm{b},p_{\mathrm{T}}^{miss})$ near $m_{\mathrm{t}}$), as this is an important feature in our signal (Figure \ref{SMprocessNewPhysics}). A trained BDT can return the discriminating power of each of its inputs: we found that the plotted kinematic variables 
(i.e., $\mathrm{p}_{\mathrm{T}}(\mathrm{b}_1)$, $\mathrm{p}_{\mathrm{T}}(\mathrm{b}_{2})$, $\Delta R(\mathrm{b}_{1}, \mathrm{b}_{2})$, and $m(\mathrm{b}_{1}, \mathrm{b}_{2})$) 
were among the most productive variables from this standpoint, producing about 60-75\% of signal significance (depending on $\mathrm{Z}^{\prime}$ mass), but the inclusion of all 47 variables does 
provide a non-trivial enhancement.

\begin{table}[]
\begin{center}
\caption {Preliminary event selection criteria used to filter events that are passed to the gradient boosting algorithm. A $\Delta R(p_{i},p_{j}) > 0.3$ requirement is applied to all the particle candidate pairs $p_{i},p_{j}$.}
\label{tab:selection_criteria}
\begin{tabular}{ l  c r}\hline\hline

  Variable &     & Threshold \\
  \hline
   $p_{\mathrm{T}}(j)$ & & $>30$ $\mathrm{GeV}$\\
   $|\eta(j)|$ & & $< 5.0$\\
  $|\eta($\textrm{b} jets$)|$ &  & $< 2.5$\\
  $p_{T}($\textrm{b} jets$)$ & & $> 30$ $\mathrm{GeV}$\\
  $N($\textrm{b} jets$)$ & & $ > 3$\\
  $N(\ell)$ & & $ = 1$\\
  $|\eta(\ell)|$ &  & $< 2.5$\\
  $p_{\mathrm{T}}(\ell)$ & & $> 25$ $\mathrm{GeV}$\\
  $\Delta R(p_{i},p_{j})$ & & $> 0.3$\\ 
 % $\Delta R(b_{i},b_{j})$ & & $> 0.3$\\
 % $\Delta R(\ell,b_{i})$ & & $> 0.3$\\

   \hline\hline
 \end{tabular}
\end{center}
\end{table}

\begin{figure}[]
    \centering
    \includegraphics[width=0.48\textwidth]{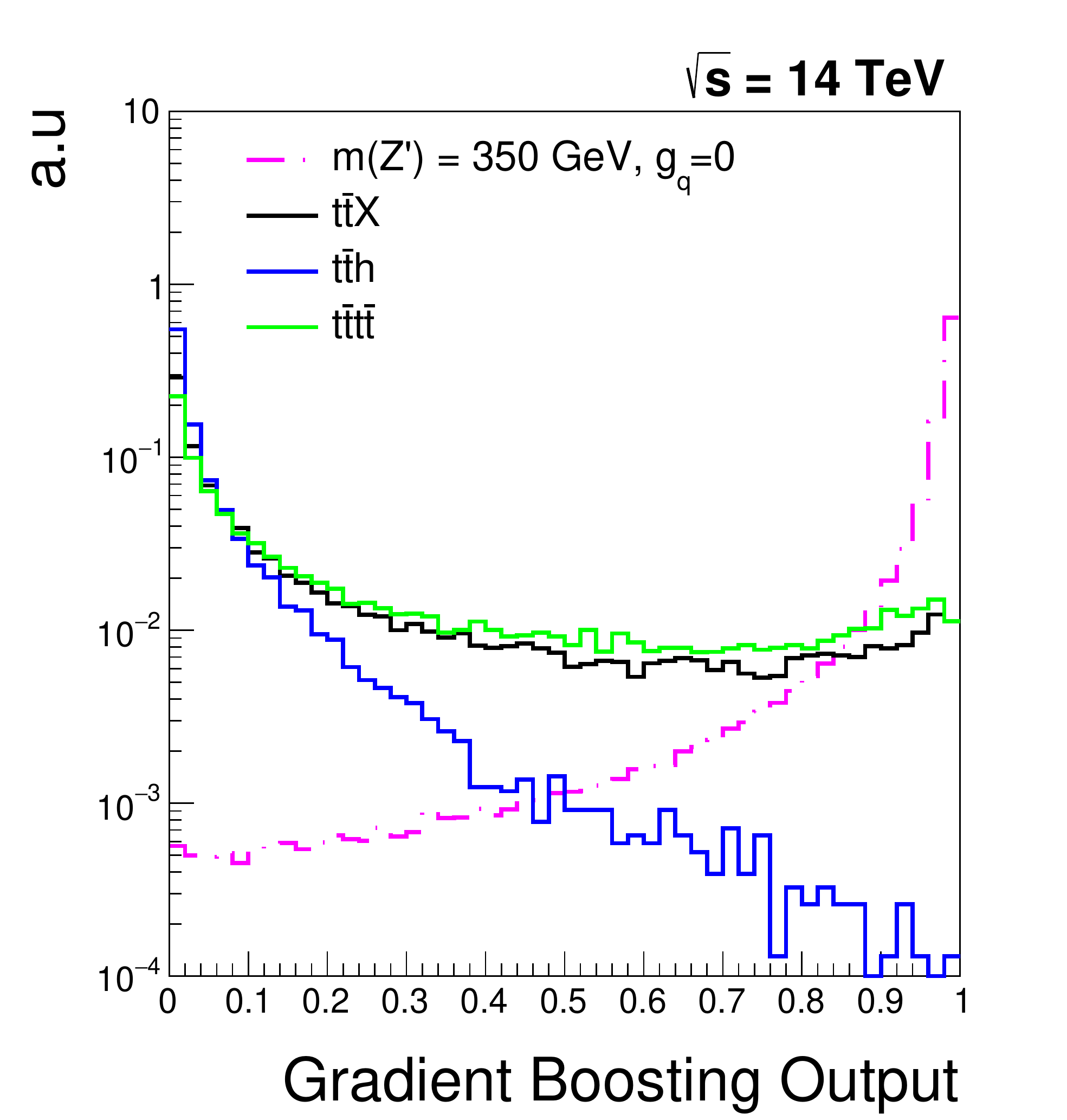}
    \caption{Output of the gradient boosting algorithm for %the %$\mathrm{Z}^{\prime} (\rightarrow \mathrm{b}\bar{\mathrm{b}})$ channel, for a signal point, 
    a $\mathrm{Z}^{\prime}$ signal with mass of 350 $\mathrm{GeV}$ and $g_{q} = 0$ coupling, and the dominant backgrounds. The distributions are normalized to unity.}
    \label{fig:mloutput350}
\end{figure}

\begin{figure}[]
    \centering
    \includegraphics[width=0.48\textwidth]{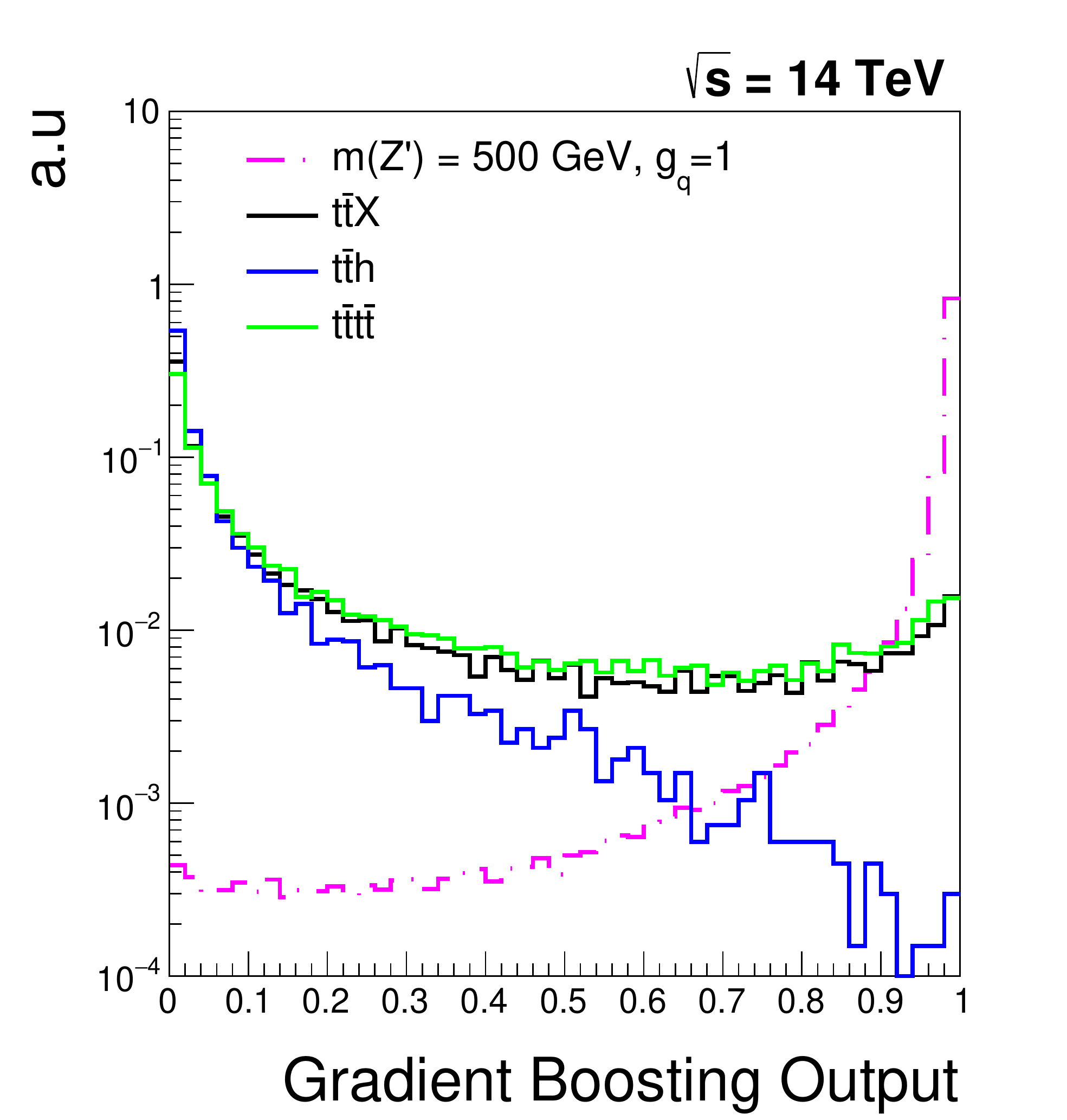}
    \caption{Output of the gradient boosting algorithm for %the %$\textrm{Z}^{\prime} (\rightarrow b\bar{b})$ channel, for a signal point, 
    a $\mathrm{Z}^{\prime}$ signal with mass of 500 $\mathrm{GeV}$ and $g_{q} = 1$ coupling, and for the most relevant backgrounds. The distributions are normalized to unity.}
    \label{fig:mloutput500}
\end{figure}

Figure \ref{fig:mloutput350} shows the distributions for the output of the BDT algorithm for a SPM1 signal benchmark point with $m(\mathrm{Z}^{\prime}) = 350$ GeV and $\{g_{q}, g_{\mathrm{Z}^{\prime}\mathrm{t}\bar{\mathrm{t}}}\} = \{0, 1\}$, and the dominant backgrounds. Figure \ref{fig:mloutput500} shows the BDT output for $m(\mathrm{Z}^{\prime}) = 500$ GeV and $\{g_{q}, g_{\mathrm{Z}^{\prime}\mathrm{t}\bar{\mathrm{t}}}\} = \{1, 1\}$. The distributions in Figures 6 and 7 are normalized to an area under the curve of unity. Table \ref{tab:eventyields14TeVm1000} shows the expected event yields per bin, normalized to cross section times luminosity times pre-selection efficiency, for a particular choice of bin ranges of the BDT output. The bins are counted from 1 to 100, going from left to right, such that bin 1 is the leftmost bin near BDT output of 0, and bin 100 is the rightmost bin near a BDT output of 1. The backgrounds dominate over the SPM1 benchmark signal yields in a large part of the BDT output spectrum, especially near zero, where the background yields are about six orders of magnitude larger. The presence of signal will be observed as an enhancement in the yields near a BDT output of unity.

\begin{table}[ht!]
    \centering
 \caption{Event yields for the main backgrounds and the signal point for $m(\textrm{Z}^{\prime}) = 1.0$ $\mathrm{TeV}$, for some of the bin entries for the output of  the gradient boosting algorithm. The events correspond to 14 $\mathrm{TeV}$, $g_{q} = 0$, and 3000 $fb^{-1}$ luminosity scenario.}
    \begin{tabular}{c|c|c|c}
    \hline
     $t\Bar{t}\; Z$ & $t\Bar{t}\; h$ & $t\Bar{t}t\Bar{t}$ & $m(\textrm{Z}^{\prime}) = 1.0$ $\mathrm{TeV}$\\
    \hline
    \multicolumn{4}{c}{Events for bin entries 1 to 10}\\
    
    \hline
    
    3.686e+06 & 7.582e+04 & 7719  & 0.110 \\
    1.088e+05 & 682.5  &  229.6  & 0.060  \\
    4.969e+04 & 230.0  &  105.4  & 0.062    \\
    2.753e+04 & 118.7  &  59.53  & 0.045 \\
    1.853e+04 & 83.45  &  42.68  & 0.043  \\
    1.258e+04 & 49.14  &  27.44  & 0.029  \\
    1.06e+04  & 39.87  &  24.52  & 0.037  \\
    7984      & 25.96  &  16.45  & 0.031      \\
    6774      & 13.91  &  12.11  & 0.035      \\
    5468      & 12.98  &  11.70  & 0.025    \\
    
    \hline
     \multicolumn{4}{c}{Events for bin entries 41 to 50}\\
    
    \hline
    870.9  & 0  &  1.412      & 0.023 \\
    822.6  & 1.855  &  1.211  & 0.019  \\
    919.3  & 0.927  &  1.211  & 0.013  \\
    435.5  & 1.855  &  1.110  & 0.014    \\
    290.3  & 1.855  &  1.513  & 0.019   \\
    629.0  & 1.855  &  1.211  & 0.019     \\
    725.8  & 0.927  &  0.908  & 0.020  \\
    338.7  & 1.855  &  0.505  & 0.020  \\
    580.6  & 1.855  &  1.009  & 0.014   \\
    387.1  & 0.0  &    0.706  & 0.030      \\
    \hline
     \multicolumn{4}{c}{Events for bin entries 91 to 100}\\
    \hline
    387.1  & 0.0  &  1.11  & 0.182  \\
    629    & 0.927  &  1.412  & 0.240  \\
    580.6  & 0.0  &  0.8071  & 0.247  \\
    725.8  & 1.855  &  1.009  & 0.335 \\
    774.2  & 1.855  &  1.11  & 0.419  \\
    387.1  & 2.782  &  1.513  & 0.589  \\
    919.3  & 1.855  &  1.614  & 0.886  \\
    629  & 0.9273  &  2.724  & 1.605  \\
    1403  & 0.927  &  3.027  & 3.754  \\
    2952  & 3.709  &  3.935  & 214.0 \\
\hline
    \end{tabular}
    \label{tab:eventyields14TeVm1000}
\end{table}

\section{Results}\label{sec:Results}

Using the BDT distributions normalized to cross section times pre-selection efficiency times luminosity, we calculate the expected experimental signal significance of the proposed search methodology, for different signal models, LHC operation conditions, and integrated luminosity scenarios. As noted earlier, we consider three values for the total integrated luminosity at the LHC: ($i$) 150 fb$^{-1}$, which is approximately the amount of $\mathrm{pp}$ data already collected by the ATLAS and CMS experiments; ($ii$) 300 fb$^{-1}$, expected in the next few years; and ($iii$) 3000 fb$^{-1}$, expected by the end of the High Luminosity LHC era. The significance is calculated using the expected bin-by-bin yields of the BDT output distribution in a profile likelihood fit, using the ROOTFit \cite{Butterworth:2015oua} package developed by CERN. Similar to Refs.~\cite{Florez:2021zoo, Florez:2019tqr, Florez:2018ojp, Florez:2017xhf, VBFZprimePaper, Florez:2016lwi}, the signal significance $Z_{sig}$ is determined using the probability of obtaining the same test statistic with the background-only hypothesis and the signal plus background hypothesis, defined as the local p-value. The value of $Z_{sig}$ corresponds to the point where the integral of a Gaussian distribution between $Z_{sig}$ and $\infty$ results in a value equal to the local p-value. 

Systematic uncertainties are incorporated into the significance calculation as nuissance parameters, using a log-normal prior for normalization and a Gaussian prior for shape related uncertainties. The systematic uncertainties are based on both experimental and theoretical constraints. A 3\% systematic uncertainty is used to account for experimental errors on the the estimation of the integrated luminosity collected by experiments. This is a reasonable and conservative choice based on Ref.~\cite{CMSPLT}. A systematic uncertainty is included due to the choice of PDF, with respect to the default set used to produce the simulated  signal and background samples. The PDF uncertainties were calculated following the PDF4LHC prescription~\cite{Butterworth:2015oua}, and results in up to 5\% systematic uncertainty, depending on the process. The effect of the chosen PDF set on the shape of the BDT output distribution is negligible. Other theoretical uncertainties were considered, such as the absence of higher-order contributions to the signal cross sections, which can alter the pre-selection efficiency and shapes of kinematic distributions which are fed into the BDT algorithm. This uncertainty is calculated by varying the renormalization and factorization scales by a factor of two with respect to the nominal value, and by considering the full change in the bin-by-bin yields of the BDT output distribution. They are found to be at most 3\% in a given bin. For experimental uncertainties related to the reconstruction and identification of bottom quarks, Ref.~\cite{CMSbtag} reports a systematic uncertainty of 1-5\%, depending on $p_{\mathrm{T}}$ and $\eta$ of the b-jet candidate. However, we assume a conservative 5\% uncertainty per b-jet candidate, independent of $p_{\mathrm{T}}$ and $\eta$, which is correlated between signal and background processes with genuine bottom quarks, and correlated across BDT bins for each process. The electron and muon reconstruction, identification, and isolation requirements have an uncertainty of 2\%, while a conservative 3\% systematic uncertainty is set on the variation of the electron and muon energy/momentum scale and resolution \cite{CMS:2022fsw, ATLAS:2022uhq}. We assumed 2-5\% jet energy scale uncertainties, depending on $\eta$ and $p_{\mathrm{T}}$, resulting in shape-based uncertainties on the BDT output distribution that range from 1\% to 4\%, depending on the BDT bin. Finally, we consider a 10\% systematic uncertainty associated with possible errors on the background predictions, which are uncorrelated between background processes.

\begin{figure}[]
    \centering
    \includegraphics[width=0.48\textwidth]{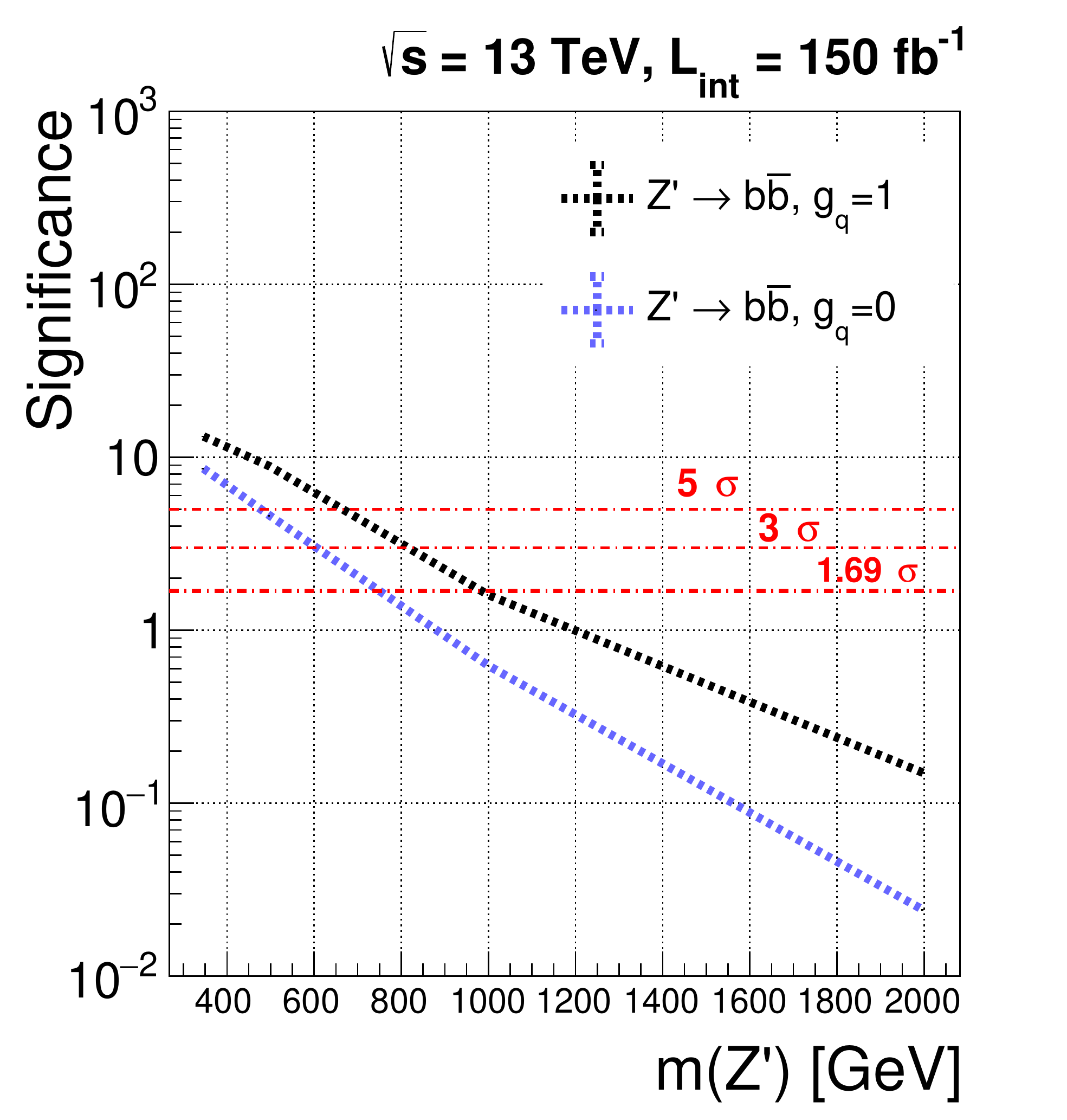}
    \caption{Expected signal significance as function of reconstructed mass, at  $\sqrt{s} = 13$ $\mathrm{TeV}$ and $150 fb^{-1}$ luminosity, for the $g_{q} = 0,1$ and $g_{\mathrm{Z}^{\prime}\mathrm{t}\bar{\mathrm{t}}} =1$ benchmark coupling scenarios. The $1.69 \sigma$ reference point for exclusion, and the $3 \sigma$ and $5 \sigma$ points for discovery sensitivity are shown as red-dashed lines.}
    \label{fig:13tevsig}
\end{figure}

\begin{figure}[]
    \centering
    \includegraphics[width=0.48\textwidth]{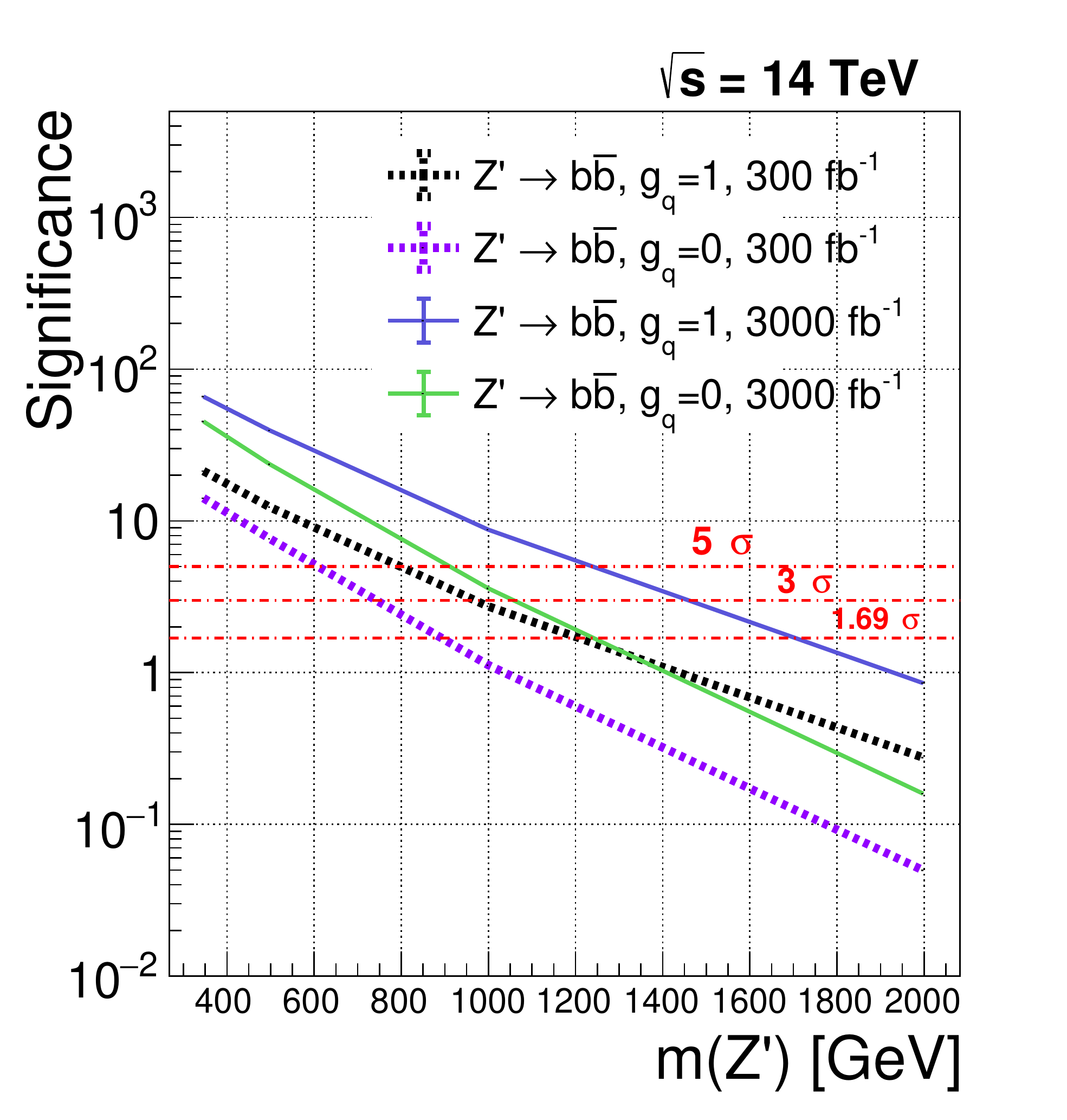}
    \caption{Expected signal significance as function of reconstructed mass, at  $\sqrt{s} = 14$ TeV and $300 fb^{-1}$ ($3000 fb^{-1}$) luminosity, for the $g_{q} = 0,1$ and 
    $g_{\mathrm{Z}^{\prime}\mathrm{t}\bar{\mathrm{t}}} =1$ benchmark
    benchmark coupling scenarios. The $1.69 \sigma$ reference point for exclusion, and the $3 \sigma$ and $5 \sigma$ points for discovery sensitivity are shown as red-dashed lines.}
    \label{fig:14tevsig}
\end{figure}

Figure \ref{fig:13tevsig} shows the SPM1 signal significance as function of $\textrm{Z}^{\prime}$ mass, for the $\{ g_{q}, g_{\mathrm{Z}^{\prime}\mathrm{t}\bar{\mathrm{t}}} \} = \{ 0,1 \}$ and $\{ g_{q}, g_{\mathrm{Z}^{\prime}\mathrm{t}\bar{\mathrm{t}}} \} = \{ 1,1 \}$ coupling scenarios, assuming $\sqrt{s} = 13$ $\mathrm{TeV}$ and 150 $\mathrm{fb}^{-1}$. A signal significance of 1.69$\sigma$ is our threshold to define expected exclusion at 95\% confidence level, while 3$\sigma$ (5$\sigma$) significance defines evidence (discovery) of new physics.  For the $\{ g_{q}, g_{\mathrm{Z}^{\prime}\mathrm{t}\bar{\mathrm{t}}} \} = \{ 1,1 \}$ scenario, the analysis shows potential to exclude masses below 1.0 TeV, and achieve greater than 3$\sigma$ (5$\sigma$) signal sensitivity for $\mathrm{Z}^{\prime}$ masses below 800 (675) GeV. For the SPM1 scenario with $\{ g_{q}, g_{\mathrm{Z}^{\prime}\mathrm{t}\bar{\mathrm{t}}} \} = \{ 0,1 \}$, the expected exclusion range is $m(\mathrm{Z}^{\prime}) < 780$ GeV, and the 3$\sigma$ (5$\sigma$) reach is $m(\mathrm{Z}^{\prime}) < 600$ (500) GeV. Figure \ref{fig:14tevsig} shows the results for the same scenarios, but considering $\mathrm{pp}$ collisions at $\sqrt{s} = 14$ $\mathrm{TeV}$ and integrated luminosities of 300 $\mathrm{fb}^{-1}$ and 3000 $\mathrm{fb}^{-1}$. For the $\{ g_{q}, g_{\mathrm{Z}^{\prime}\mathrm{t}\bar{\mathrm{t}}} \} = \{ 1,1 \}$ scenario and assuming an integrated luminosity of 3000 $\mathrm{fb}^{-1}$, the expected exclusion bound goes up to $m(\mathrm{Z}^{\prime}) < 1.7$ $\mathrm{TeV}$, while the 3$\sigma$ reach improves to $m(\mathrm{Z}^{\prime}) < 1.45$ $\mathrm{TeV}$.

\begin{table}[ht!]
    \centering
 \caption{Projected signal significance for our second simplified model, considering the $c_{\mathrm{t}} = 1$ coupling scenario with varying $\mathrm{Z}^{\prime}$ masses. The calculations are performed at $\sqrt{s} = 14$ $\mathrm{TeV}$ and assuming both $300$ $\mathrm{fb}^{-1}$ and $3000$ $\mathrm{fb}^{-1}$.}
    \begin{tabular}{c|c|c}
    \hline
    %\hline
    \multicolumn{1}{c|}{$m(\mathrm{Z}^{\prime})$}&\multicolumn{1}{c|}{300 fb $^{-1}$}&\multicolumn{1}{c}{3000 fb$^{-1}$}\\

    \hline

    250 GeV  & 59.7   & 188.8\\
    350 GeV  & 45.1   & 142.8\\
    500 GeV  & 23.8   & 75.4\\
    1000 GeV & 3.31   & 10.48\\
    1500 GeV & 1.68   & 5.41\\
    2000 GeV &0.135   & 0.427\\
    
    \hline
    \hline
    \end{tabular}
    \label{tabla_topphilic}
\end{table}

\begin{figure}[!t]
    \centering
    \includegraphics[width=0.48\textwidth]{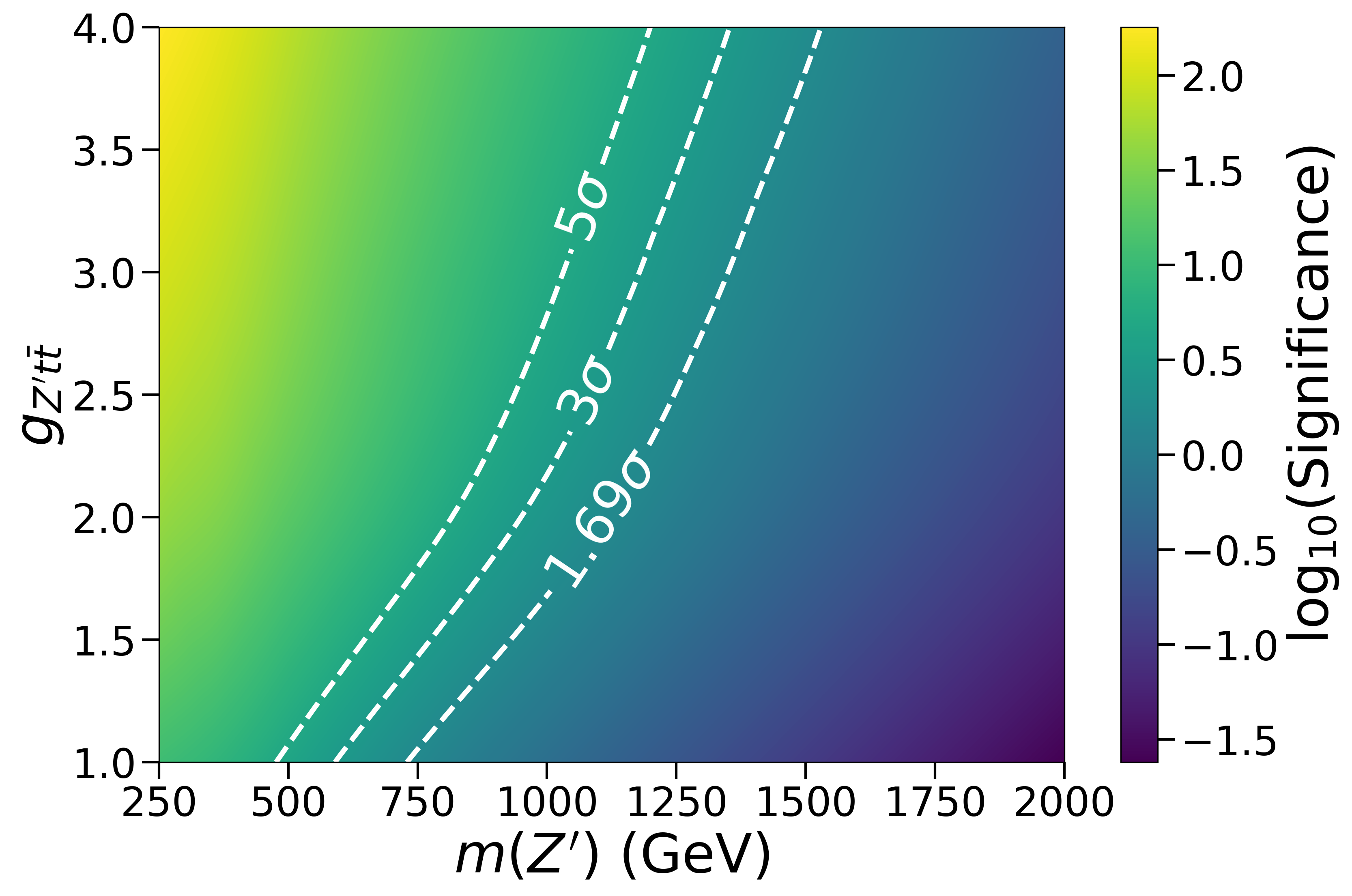}
    \caption{Projected signal significance for the $g_{q} = 0$ benchmark model for different $g_{tt}$ coupling scenarios and $\textrm{Z}^{\prime}$ masses. The estimates are performed at $\sqrt{s} = 13$ TeV and $150 fb^{-1}$.}
    \label{fig:sig13tev150bbg0}
\end{figure}

\begin{figure}[!t]
    \centering
    \includegraphics[width=0.48\textwidth]{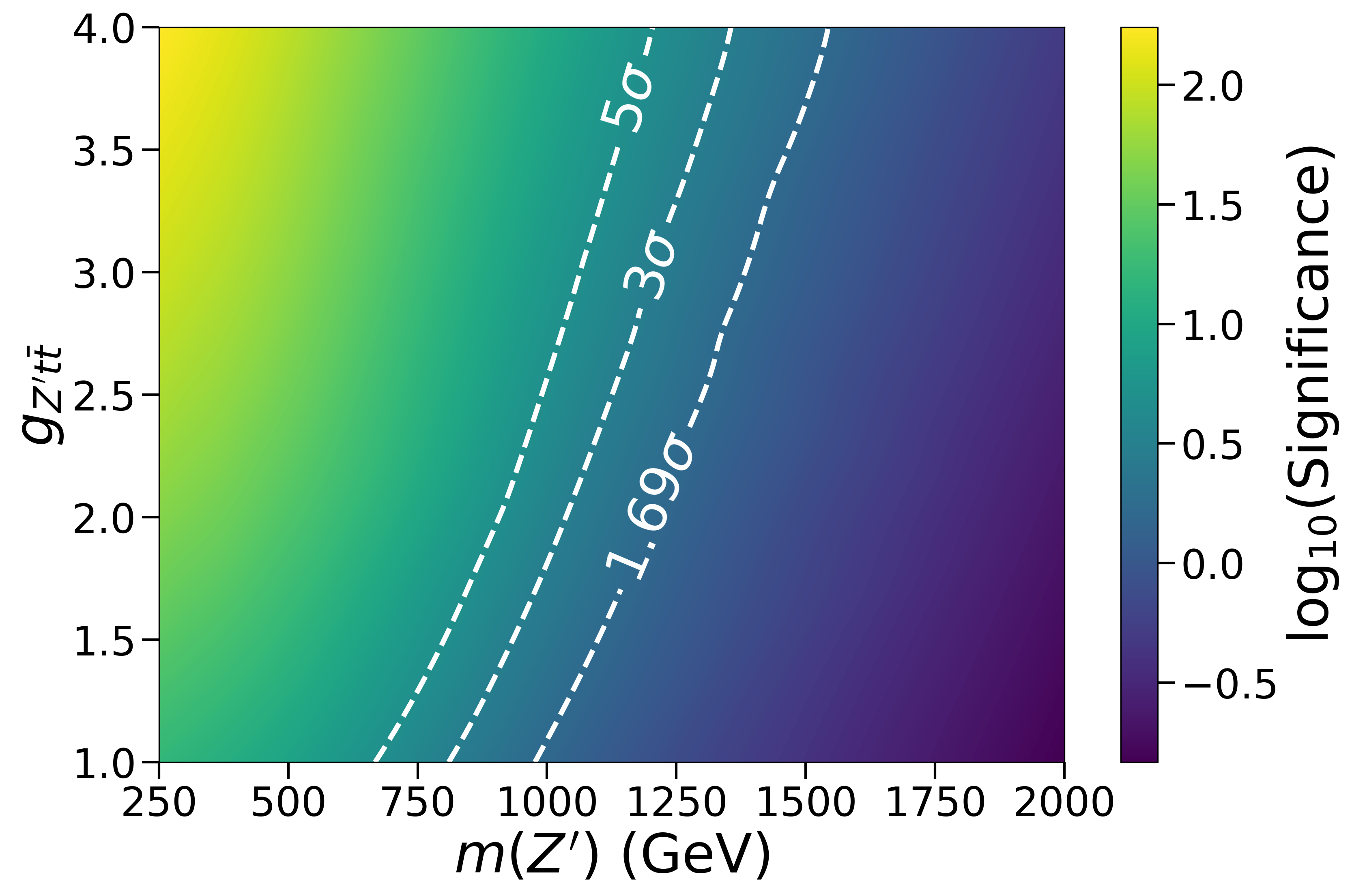}
    \caption{Projected signal significance for the $g_{q} = 1$ benchmark model for different $g_{tt}$ coupling scenarios and $\textrm{Z}^{\prime}$ masses. The estimates are performed at $\sqrt{s} = 13$ $\mathrm{TeV}$ and $150 fb^{-1}$.}
    \label{fig:sig13tev150bbg1}
\end{figure}

\begin{figure}[!t]
    \centering
    \includegraphics[width=0.48\textwidth]{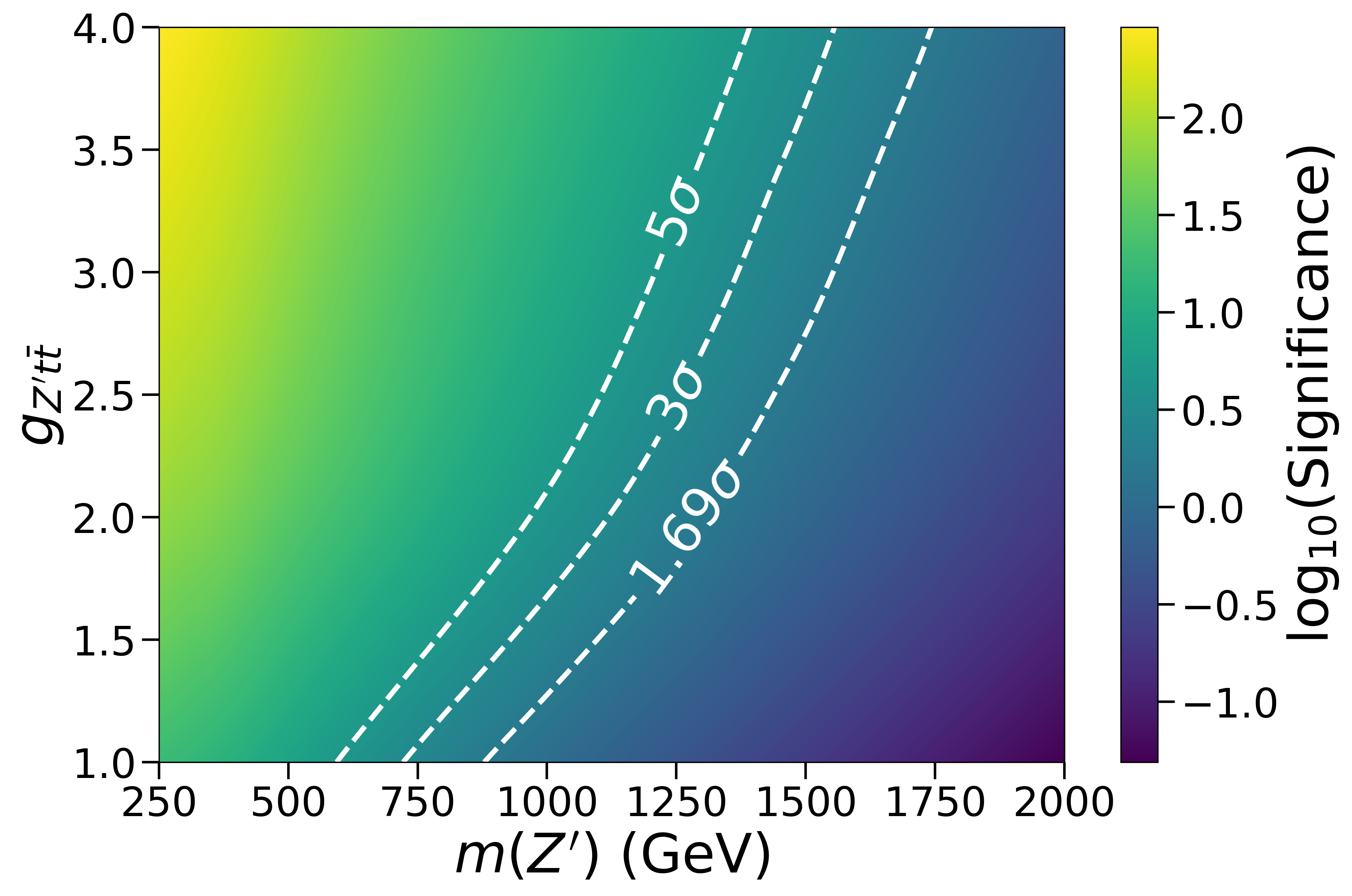}
    \caption{Projected signal significance for the $g_{q} = 0$ benchmark model for different $g_{tt}$ coupling scenarios and $\textrm{Z}^{\prime}$ masses. The estimates are performed at $\sqrt{s} = 14$ $\mathrm{TeV}$ and $300 fb^{-1}$.}
    \label{fig:sig14tev300bbg0}
\end{figure}

\begin{figure}[!t]
    \centering
    \includegraphics[width=0.48\textwidth]{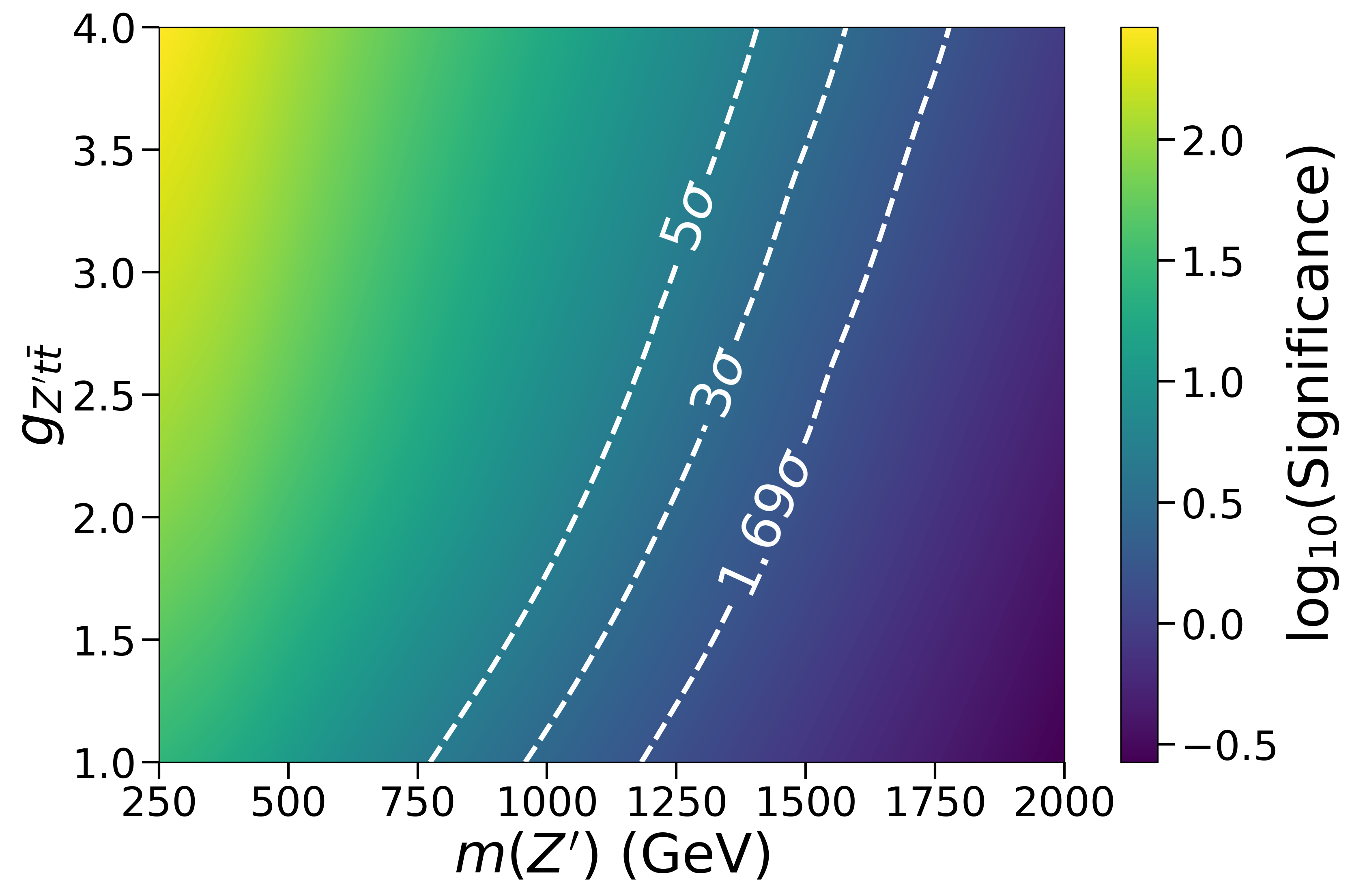}
    \caption{Projected signal significance for the $g_{q} = 1$ benchmark model for different $g_{tt}$ coupling scenarios and $\textrm{Z}^{\prime}$ masses. The estimates are performed at $\sqrt{s} = 14$ $\mathrm{TeV}$ and $300 fb^{-1}$.}
    \label{fig:sig14tev300bbg1}
\end{figure}

\begin{figure}[!t]
    \centering
    \includegraphics[width=0.48\textwidth]{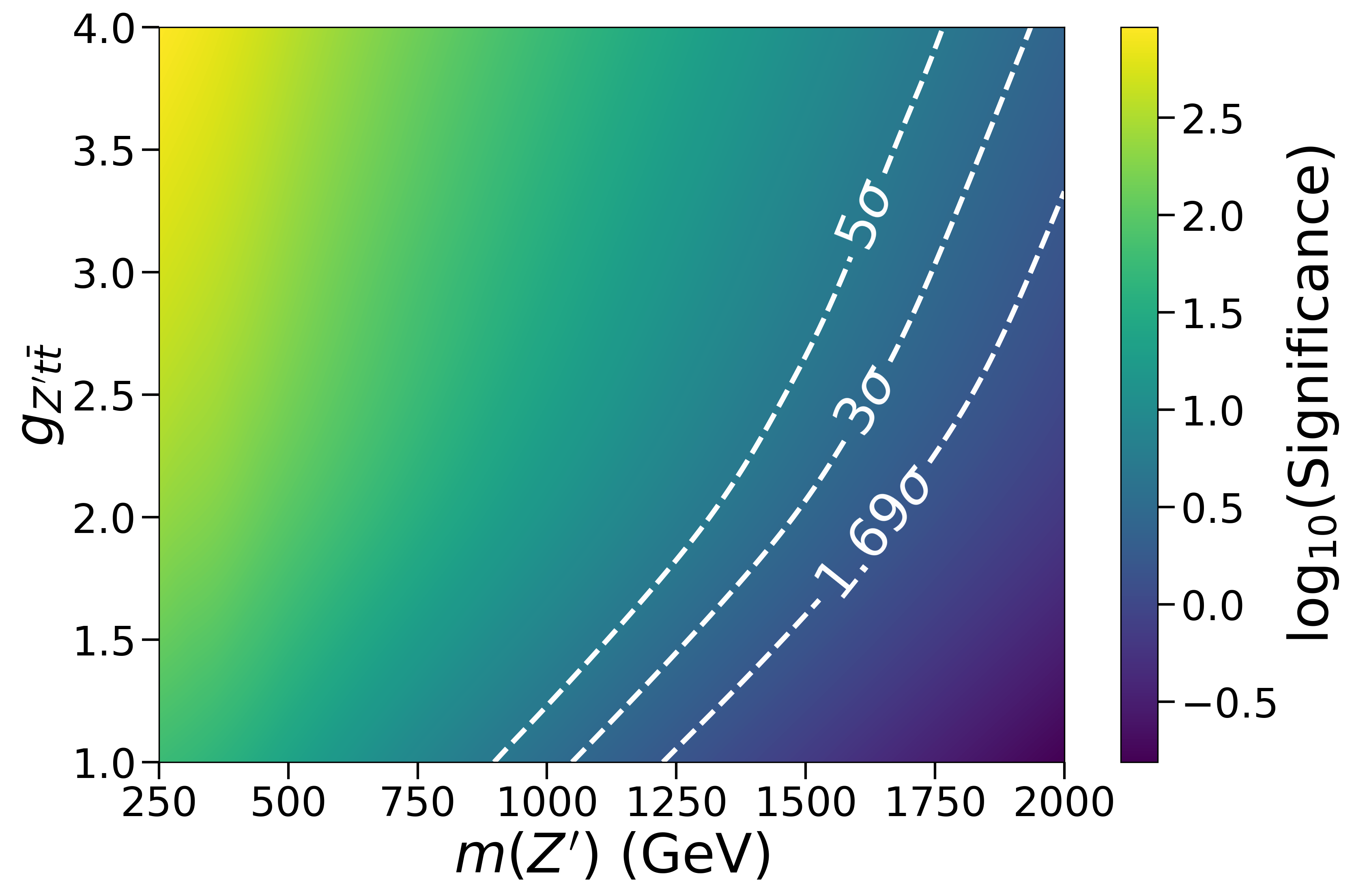}
    \caption{Projected signal significance for the $g_{q} = 0$ benchmark model for different $g_{tt}$ coupling scenarios and $\textrm{Z}^{\prime}$ masses. The estimates are performed at $\sqrt{s} = 14$ $\mathrm{TeV}$ and $3000 fb^{-1}$.}
    \label{fig:sig14tev3000bbg0}
\end{figure}

\begin{figure}[!t]
    \centering
    \includegraphics[width=0.48\textwidth]{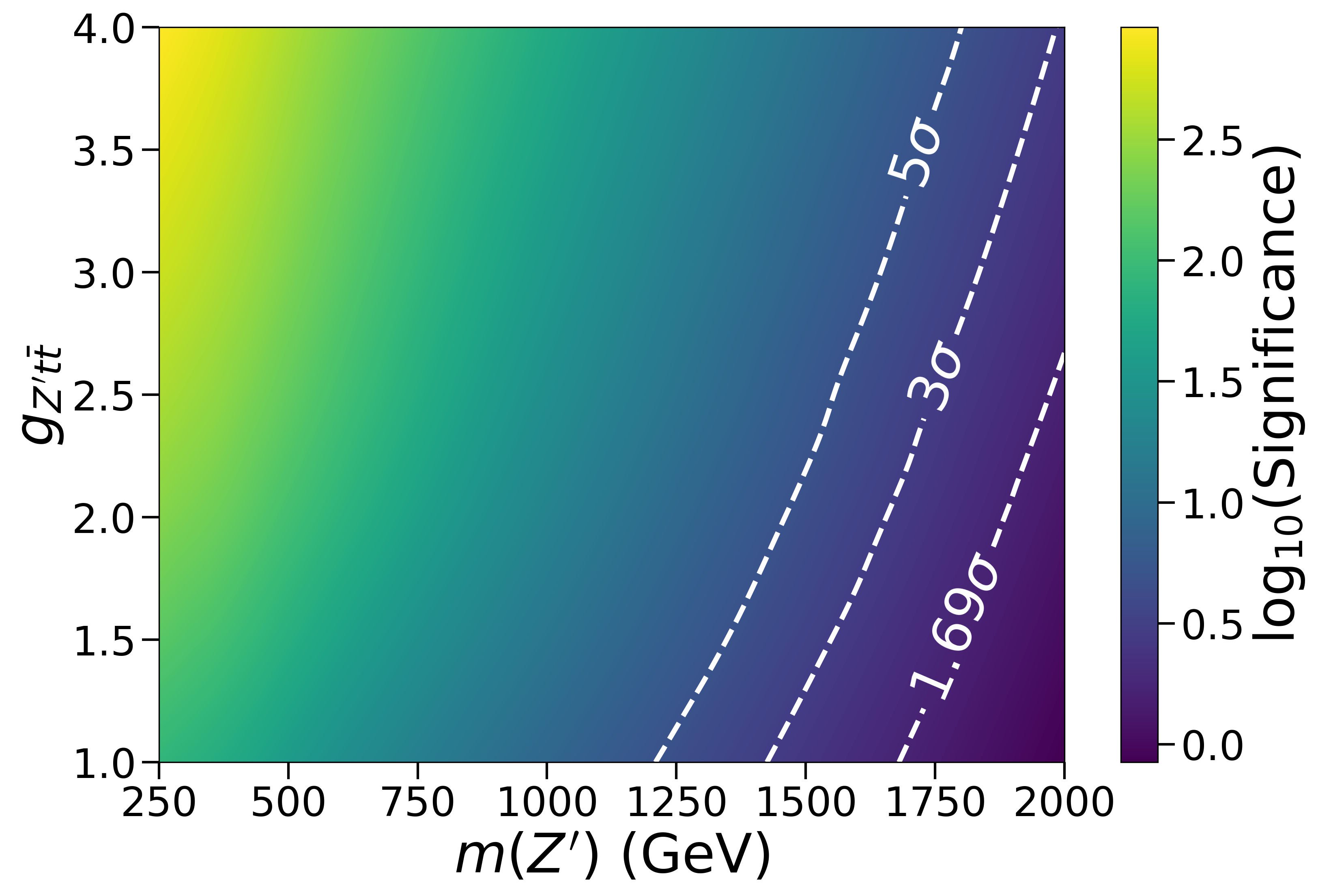}
    \caption{Projected signal significance for the $g_{q} = 1$ benchmark model for different $g_{tt}$ coupling scenarios and $\textrm{Z}^{\prime}$ masses. The estimates are performed at $\sqrt{s} = 14$ $\mathrm{TeV}$ and $3000 fb^{-1}$.}
    \label{fig:sig14tevbb3000g1}
\end{figure}

We also estimate the expected signal significance for different SPM1 coupling scenarios of the $\mathrm{Z}^{\prime}$ boson to $\mathrm{t}$/$\mathrm{b}$ quarks. Figure \ref{fig:sig13tev150bbg0} shows the signal significance for different $g_{\mathrm{Z}^{\prime}\mathrm{t\bar{t}}}$ and $m(\mathrm{Z}^{\prime})$ scenarios, with suppressed couplings to first and second generation quarks ($g_{q} = 0$), assuming $\sqrt{s} = 13$ $\mathrm{TeV}$ and 150 $\mathrm{fb}^{-1}$. Figure \ref{fig:sig13tev150bbg1} shows the corresponding results for the same $\{ g_{\mathrm{Z}^{\prime}\mathrm{t\bar{t}}}, m(\mathrm{Z}^{\prime}) \}$ combinations, but using $g_{q} = 1$. The results for $\sqrt{s} = 14$ $\mathrm{TeV}$, assuming 300 $\mathrm{fb}^{-1}$ and 3000 $\mathrm{fb}^{-1}$, are presented in
Figures~\ref{fig:sig14tev300bbg0}--\ref{fig:sig14tevbb3000g1} for different $\{ g_{q},  g_{\mathrm{Z}^{\prime}\mathrm{t\bar{t}}}, m(\mathrm{Z}^{\prime}) \}$ combinations.

Table \ref{tabla_topphilic} shows the SPM2 signal significance as function of $m(\textrm{Z}^{\prime})$ and integrated luminosity, for the $\{ c_{\mathrm{t}}, \theta \} = \{ 1,\pi/2 \}$ scenario, assuming $\sqrt{s} = 14$ $\mathrm{TeV}$. The expected SPM2 exclusion range is $m(\mathrm{Z}^{\prime}) < 1.5$ TeV at $\mathrm{L} = 300$ fb$^{-1}$, while the 5$\sigma$ discovery reach is $m(\mathrm{Z}^{\prime}) < 1.5$ TeV for the 3000 fb$^{-1}$ expected by the end of the high luminosity LHC era.

\section{Discussion}
As the LHC continues to run with pp collisions at the highest energy, and with the slow increase in luminosity expected of the high-luminosity program of the accelerator, it is an important matter to ponder why certain searches for new physics have not provided strong evidence for discovery, and consider unexplored possibilities. In this work, we examine the phenomonology of a vector like $\mathrm{Z}^{\prime}$ boson favoring higher-generation fermions (anogenophilic), in particular coupling to third generation fermions (tritogenophilic). This scenario is well motivated and arises in many theories that extend the SM~\cite{Hill:1994hp,Randall:1999ee, Randall:1999vf,Davoudiasl:2000wi,Branco:2011iw,Gori:2016zto,BhupalDev:2014bir,Gunion:1989we,left-right,Dobrescu:2007yp,Chen:2008hh,Bai:2010dj,Berger:2010fy,Zhang:2010kr,Gerbush:2007fe,ArkaniHamed:2002qy, Pomarol:2008bh,Matsedonskyi:2012ym,Gripaios:2014pqa,Liu:2015hxi,Cacciapaglia:2015eqa,Barbieri:2007bh,Panico:2011pw,DeCurtis:2011yx,Marzocca:2012zn,Bellazzini:2012tv,Panico:2012uw,Bellazzini:2014yua,Haisch:2015ioa,Buckley:2014fba,Cox:2015afa,Zhang:2012da,Baek:2016lnv,Arina:2016cqj,DHondt:2015nat,Baek:2017sew}. It also seems to appear as a possible, although not yet confirmed, pattern in precision measurements of the $B$-physics sector~\cite{Lees:2012xj,Lees:2013uzd,Huschle:2015rga,Sato:2016svk,Hirose:2016wfn,Hirose:2017dxl, Belle:2019rba, Aaij:2015yra,Aaij:2017uff,Aaij:2017tyk,Wehle:2016yoi,Aaij:2013qta,Aaij:2014pli,Aaij:2014ora,Aaij:2015oid,Aaij:2015esa,Aaij:2017vbb,Aaij:2019wad} and the measurement of the muon anomalous magnetic moment~\cite{Muong-2:2021ojo}. An anogenophilic $\mathrm{Z}^{\prime}$ has already been investigated phenomenologically or experimentally for the case in which the new boson is produced in association with two top quarks and decays to two top quarks (top-philic~\cite{Greiner:2014qna, Kim:2016plm, Fox:2018ldq}), tau/muon leptons (~\cite{Abdullah:2019dpu,Kamenik:2017tnu}), or muon/electron leptons (~\cite{CMS:2019lwf}). Here we have presented a feasibility study for the $\mathrm{Z}^{\prime}$ decay into two $\mathrm{b}$ quarks. 
The study has been performed under the context of $\mathrm{pp}$ collisions at the LHC, at $\sqrt{s} = 13$ $\mathrm{TeV}$ and $\sqrt{s} = 14$ $\mathrm{TeV}$, using a BDT algorithm to optimize the signal to background separation and maximize exclusion or discovery potential. Various coupling scenarios for the $\mathrm{Z}^{\prime}$ have been considered, including suppressed couplings to light flavour quarks ($g_{q} = 0$), enhanced couplings to third generation fermions, and preferential couplings to top and bottom quarks ($g_{\mathrm{Z}^{\prime}t\bar{t}}$). Under the SPM1 $g_{q} = 1$ ($g_{q} = 0$) scenario, at $\sqrt{s} = 13$ $\mathrm{TeV}$ and integrated luminosity of 150 $\mathrm{fb}^{-1}$, $\mathrm{Z}^{\prime}$ masses up to 1.0 $\mathrm{TeV}$ (780 $\mathrm{GeV}$) can be excluded at 95\% confidence level, while 5$\sigma$ discovery potential exists for masses below 675 $\mathrm{GeV}$ (500 $\mathrm{GeV}$). For the high luminosity era of the LHC with $\sqrt{s} = 14$ $\mathrm{TeV}$ and integrated luminosity of 3000 $\mathrm{fb}^{-1}$, $\mathrm{Z}^{\prime}$ masses up to 1.70 $\mathrm{TeV}$ (1.25 $\mathrm{TeV}$) can be excluded for the SPM1 $g_{q} = 1$ ($g_{q} = 0$) scenario, while the 5$\sigma$ discovery reach is $m(\mathrm{Z}^{\prime}) < 1.25$ $\mathrm{TeV}$ (900 $\mathrm{GeV}$). For the SPM2 benchmark scenario with $c_{\mathrm{t}} = 1$ and $\theta = \pi / 2$, the discovery (exclusion) reach is 1.5 (1.7) TeV at $\sqrt{s} = 14$ $\mathrm{TeV}$ and integrated luminosity of 3000 $\mathrm{fb}^{-1}$. As noted previously, the projected sensitivity using the SPM2 scenario serves as a good comparison with other search strategies. For example, the authors of Ref.~\cite{Kim:2016plm} examined the high luminosity LHC sensitivity to these  anogenophilic scenarios using the $\mathrm{pp}\to\mathrm{t\bar{t}}\mathrm{Z}^{\prime}\to\mathrm{t\bar{t}}\mathrm{t\bar{t}}$ final state with boosted top tagging algorithms, and reported a projected $2\sigma$ reach of approximately $m(\mathrm{Z}^{\prime}) < 1.5$ TeV for the same coupling scenario of $c_{\mathrm{t}}=1$, assuming an integrated luminosity of 3000 fb$^{-1}$. That result is to be compared with the stronger projected significance of $> 5.41\sigma$ for $m(\mathrm{Z}^{\prime}) < 1.5$ TeV in Table~\ref{tabla_topphilic}, using the strategy presented in this paper. Additionally, Ref.~\cite{Kim:2016plm} reports that a $> 5\sigma$ discovery reach is attainable for $m(\mathrm{Z}^{\prime}) = 1.5$ TeV if $c_{\mathrm{t}} > 1.65$. For comparison, Table~\ref{tabla_topphilic} already shows a significance of $5.41\sigma$ for $m(\mathrm{Z}^{\prime}) = 1.5$ TeV with a smaller coupling of $c_{\mathrm{t}} = 1$. We also point out that these comparisons are conservative since the studies outlined in Ref.~\cite{Kim:2016plm} assume a 100\% branching ratio of $\mathrm{Z}^{\prime} \to \mathrm{t\bar{t}}$, which would not be the case if $\mathrm{Z}^{\prime}$ couples to both top and bottom quarks.

The main result of this paper is that probing heavy neutral gauge bosons produced in association with spectator top quarks, and decaying to a 
pair of bottom quarks, can be a key search methodology. It represents the most important anogenophilic/tritogenophilic mode for discovery at 
$m(\mathrm{Z}^{\prime}) < 2 m_{\mathrm{t}}$ where the $\mathrm{Z}^{\prime}\to \mathrm{t\bar{t}}$ decay is kinematically forbidden, and remains competitive with the $\mathrm{Z}^{\prime}\to \mathrm{t\bar{t}}$ decay mode at $\mathrm{TeV}$ scale masses, benefiting from the possibility to reconstruct the $\mathrm{Z}^{\prime}$ mass from the two highest-$p_{\mathrm{T}}$ $\mathrm{b}$ jets and resulting in events with reduced jet multiplicity. Furthermore, even if a $\mathrm{Z}^{\prime}$ boson is discovered in other search channels when $m(\mathrm{Z}^{\prime})$ is 
large, a $\mathrm{t\bar{t}}\mathrm{Z}^{\prime} \to \mathrm{t\bar{t}} \mathrm{b\bar{b}}$ search remains a key part of the search program at the LHC in order to establish the couplings of the $\mathrm{Z}^{\prime}$ to all fermions. In particular, whereas a $\mathrm{t\bar{t}}\mathrm{Z}^{\prime} \to \mathrm{t\bar{t}} \mathrm{t\bar{t}}$ search can measure the $\mathrm{Z}^{\prime}$ mass and coupling to top quarks, the proposed $\mathrm{t\bar{t}}\mathrm{Z}^{\prime} \to \mathrm{t\bar{t}} \mathrm{b\bar{b}}$ search can additionally measure the $\mathrm{Z}^{\prime}$ coupling to bottom quarks. 

The proposed data analysis represents a competitive alternative 
to complement searches already being conducted at the LHC. Those searches are based on the analysis of the mass distribution of two $\mathrm{b}$-quark jets, in the resolved or boosted regime, using events whose triggers require high-$p_{\mathrm{T}}$ jets~\cite{ATLAS:2019fgd, CMS:2022zoc, CMS:2018pwl}, $\mathrm{b}$-quark jets~\cite{CMS:2018kcg}, or a photon~\cite{ATLAS:2019itm}. In the analysis strategy considered here instead, we can rely on the presence of an electron or muon lepton originating from the decay of a spectator top, which allows an unbiased selection of $\mathrm{b}$-quark jets originating from the $\mathrm{Z}^{\prime}$, or on the possibility 
to define a trigger using both a light lepton and jets, in order to select particles with lower energy. 

Because of the above reasons, we deem that that the proposed analysis strategy should be considered in future $\mathrm{Z}^{\prime}$ searches at the LHC, by both the ATLAS and the CMS collaboration.

\section{Acknowledgements}
We thank the constant and enduring financial support received for this project from the faculty of science at Universidad de Los Andes (Bogot\'a, Colombia), the Physics \& Astronomy department at Vanderbilt University and the US National Science Foundation. This work is supported in part by NSF Award PHY-1945366 and a Vanderbilt Seeding Success Grant.

% E. References Cited
%\newpage
%\newline 
\newpage 
%\newline 
\newpage 
%\newsection{}
%\renewcommand\refname{References Cited}
\bibliography{ttbarZprime} % Entries are in the "references.bib" file
% I prefer to use the IEEE bibliography style. 
% That's  NOT required by the NSF guidelines. 
% Feel Free to use whatever style you prefer
\bibliographystyle{IEEEtran}%See 

%https://github.com/ionaic/bibtex-ieeetran-urldate/blob/master/IEEEtranUrldate.bst for all options
%https://www.overleaf.com/learn/latex/Bibliography_management_in_LaTeX
%\bibliographystyle{alpha} 

\end{document}